\documentclass[acmtog,authorversion]{acmart}
\acmSubmissionID{511}

\usepackage[ruled]{algorithm2e}

\SetAlFnt{\small}
\SetAlCapFnt{\small}
\SetAlCapNameFnt{\small}
\SetAlCapHSkip{0pt}

\setcopyright{rightsretained}
\acmJournal{TOG}
\acmYear{2021}\acmVolume{40}\acmNumber{4}\acmArticle{59}\acmMonth{8}
\acmDOI{10.1145/3450626.3459863}

\usepackage{listings}
\usepackage{epsfig}
\usepackage{graphicx}
\usepackage{amsmath}
\usepackage{color}
\usepackage{colortbl}

\usepackage[tableposition=top]{caption}

\DeclareMathOperator*{\argmin}{arg\,min}

\begin{document}

\title{Mixture of Volumetric Primitives for Efficient Neural Rendering}

\author{Stephen Lombardi}
\affiliation{%
  \institution{Facebook Reality Labs}
  \country{USA}
}
\email{stephen.lombardi@fb.com}

\author{Tomas Simon}
\affiliation{%
  \institution{Facebook Reality Labs}
  \country{USA}
}
\email{tsimon@fb.com}

\author{Gabriel Schwartz}
\affiliation{%
  \institution{Facebook Reality Labs}
  \country{USA}
}
\email{gbschwartz@fb.com}

\author{Michael Zollhoefer}
\affiliation{%
  \institution{Facebook Reality Labs}
  \country{USA}
}
\email{zollhoefer@fb.com}

\author{Yaser Sheikh}
\affiliation{%
  \institution{Facebook Reality Labs}
  \country{USA}
}
\email{yasers@fb.com}

\author{Jason Saragih}
\affiliation{%
  \institution{Facebook Reality Labs}
  \country{USA}
}
\email{jsaragih@fb.com}



\begin{teaserfigure}
    \centering
    \includegraphics[width=\linewidth]{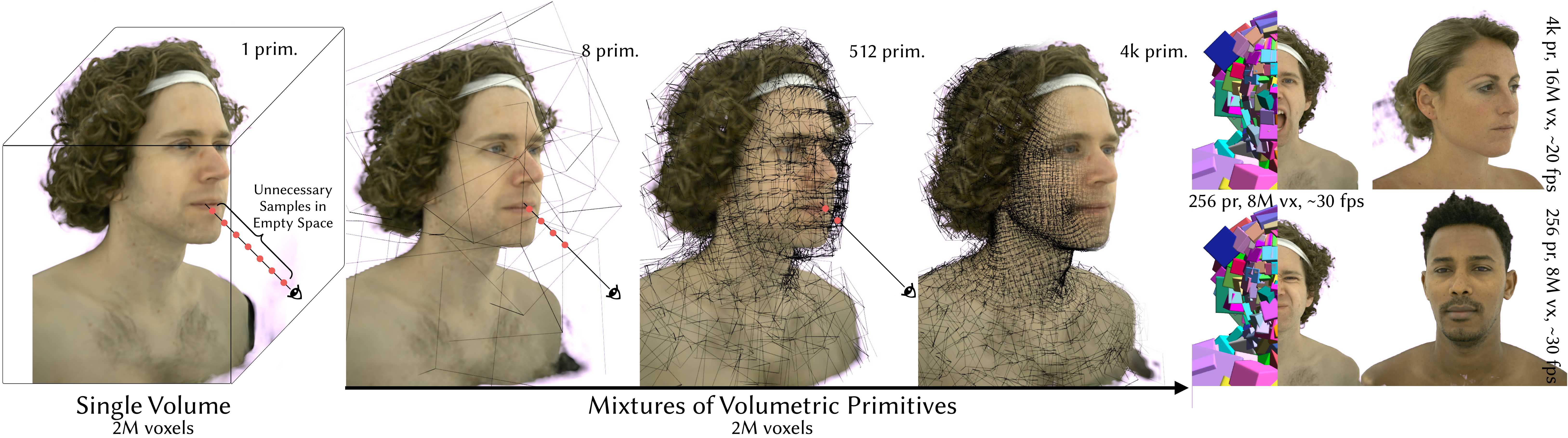}
    \vspace{-0.6cm}
    \caption{
    Mixture of Volumetric Primitives (MVP) inherits the strengths of both volumetric as well as primitive-based approaches, while avoiding many of their pitfalls.
    It enables state-of-the-art results for rendering dynamic objects in terms of rendering quality and runtime performance.
    }
    \label{fig:MVP}
\end{teaserfigure}

\begin{abstract}
Real-time rendering and animation of humans is a core function in games, movies, and telepresence applications.
Existing methods have a number of drawbacks we aim to address with our work. Triangle meshes have difficulty modeling thin structures like hair, volumetric representations like Neural Volumes are too low-resolution given a reasonable memory budget, and high-resolution implicit representations like Neural Radiance Fields are too slow for use in real-time applications.
We present Mixture of Volumetric Primitives (MVP), a representation for rendering dynamic 3D content that combines the completeness of volumetric representations with the efficiency of primitive-based rendering, e.g., point-based or mesh-based methods.
Our approach achieves this by leveraging spatially shared computation with a convolutional architecture and by minimizing computation in empty regions of space with volumetric primitives that can move to cover only occupied regions.
Our parameterization supports the integration of correspondence and tracking constraints, while being robust to areas where classical tracking fails, such as around thin or translucent structures and areas with large topological variability.
MVP is a hybrid that generalizes both volumetric and primitive-based representations.
Through a series of extensive experiments we demonstrate that it inherits the strengths of each, while avoiding many of their limitations.
We also compare our approach to several state-of-the-art methods and demonstrate that MVP produces superior results in terms of quality and runtime performance.
\end{abstract}

\maketitle

\section{Introduction}
Photo-realistic rendering of dynamic 3D objects and scenes from 2D image data is a central focus of research in computer vision and graphics.
Volumetric representations have seen a resurgence of interest in the graphics community in recent years, driven by impressive empirical results attained using learning-based methods~\cite{lombardi2019nv,mildenhall2020nerf}.
Through the use of generic function approximators, such as deep neural networks, these methods achieve compelling results by supervising directly on raw image pixels.
Thus, they avoid the often difficult task of assigning geometric and radiometric properties, which is typically required by classical physics-inspired representations.
Leveraging the inherent simplicity of volumetric models, much work has been dedicated to extending the approach for modeling small motions~\cite{park2020nerfies}, illumination variation~\cite{Srinivasan20arxiv_NeRV}, reducing data requirements~\cite{Trevithick20arxiv_GRF,Yu20arxiv_pixelNeRF}, and learning efficiency~\cite{Tancik20arxiv_meta}.
All these methods employ a soft volumetric representation of 3D space that helps them model thin structures and semi-transparent objects realistically.

Despite the aforementioned advances in volumetric models, they still have to make a trade-off; either they have a large memory footprint or they are computationally expensive to render.
The large memory footprint drastically limits the resolution at which these approaches can operate and results in a lack of high-frequency detail.
In addition, their high computational cost limits applicability to real-time applications, such as VR telepresence \cite{wei2019VR,Escolano2016Holo}.
The ideal representation would be memory efficient, can be rendered fast, is drivable, and has high rendering quality.

Neural Volumes \cite{lombardi2019nv} is a method for learning, rendering, and driving dynamic objects captured using an outside-in camera rig.
The method is suited to objects rather than scenes as a uniform voxel grid is used to model the scene.
This grid's $O(n^3)$ memory requirement prevents the use of high resolutions, even on high-end graphics cards.
Since much of the scene is often comprised of empty space, Neural Volumes employs a warp field to maximize the utility of its available resolution.
The efficacy of this, however, is limited by the resolution of the warp and the ability of the network to learn complex inverse-warps in an unsupervised fashion.

Neural Radiance Fields (NeRF) \cite{mildenhall2020nerf} addresses the issue of resolution using a compact representation.
NeRF only handles static scenes.
Another challenge is runtime, since a multi-layer perceptron (MLP) has to be evaluated at every sample point along the camera rays.
This leads to billions of MLP evaluations to synthesize a single high-resolution image, resulting in extremely slow render times around thirty seconds per frame.
Efforts to mitigate this rely on coarse-to-fine greedy selection that can miss small structures~\cite{liu2020nsvf}.
This approach can not easily be extended to dynamics, since it relies on a static acceleration structure.

In this work, we present Mixture of Volumetric Primitives (MVP), an approach designed to directly address the memory and compute limitations of existing volumetric methods, while maintaining their desirable properties of completeness and direct image-based supervision.
It is comprised of a mixture of jointly-generated overlapping volumetric primitives that are selectively ray-marched, see Fig.~\ref{fig:MVP}.
MVP leverages the conditional computation of ray-tracing to eliminate computation in empty regions of space.
The generation of the volumetric primitives that occupy non-empty space leverages the shared computation properties of convolutional deep neural networks, which avoids the wasteful re-computation of common intermediate features for nearby areas, a common limitation of recent methods~\cite{mildenhall2020nerf,liu2020nsvf}.
Our approach can naturally leverage correspondences or tracking results defined previously by opportunistically linking the estimated placement of these primitives to the tracking results. This results in good motion interpolation.
Moreover, through a user-defined granularity parameter, MVP generalizes volumetric~\cite{lombardi2019nv} on one end, and primitive-based methods~\cite{lombardi2018DAM,aliev2019neuralpoint} on the other, enabling the practitioner to trade-off resolution for completeness in a straightforward manner.
We demonstrate that our approach produces higher quality, more driveable models, and can be evaluated more quickly than the state of the art.
Our key technical contributions are:
\begin{itemize}
    \setlength\itemsep{0.5em}
    \item A novel volumetric representation based on a mixture of volumetric primitives that combines the advantages of volumetric and primitive-based approaches, thus leading to high performance decoding and efficient rendering.
    \item A novel motion model for voxel grids that better captures scene motion, minimization of primitive overlap to increase the representational power, and minimization of primitive size to better model and exploit free space.
    \item A highly efficient, data-parallel implementation that enables faster training and real-time rendering of the learned models.
\end{itemize}

\section{Related Work}
In the following, we discuss different scene representations for neural rendering.
For an extensive discussion of neural rendering applications, we refer to \citet{tewari2020neuralrendering}.

\paragraph{Point-based Representations}
The simplest geometric primitive are points.
Point-based representations can handle topological changes well, since no connectivity has to be enforced between the points.
Differentiable point-based rendering has been extensively employed in the deep learning community to model the geometry of objects \cite{insafutdinov2018pointclouds,roveri2018network,lin2018learning,yifan2019differentiable}.
Differentiable Surface Splatting \cite{yifan2019differentiable} represents the points as discs with a position and orientation.
Lin et~al.~\cite{lin2018learning} learns efficient point cloud generation for dense 3D object reconstruction.
Besides geometry, point-based representations have also been employed extensively to model scene appearance \cite{meshry2019neuralrerendering, aliev2019neuralpoint, wiles2020synsin, lassner2020pulsar, kolos2020transpr}.
One of the drawbacks of point-based representations is that there might be holes between the points after projection to screen space.
Thus, all of these techniques often employ a network in screen-space, e.g., a U-Net \cite{ronneberger2015unet}, to in-paint the gaps.
SynSin~\cite{wiles2020synsin} lifts per-pixel features from a source image onto a 3D point cloud that can be explicitly projected to the target view.
The resulting feature map is converted to a realistic image using a screen-space network.
While the screen-space network is able to plausibly fill in the holes, point-based methods often suffer from temporal instabilities due to this screen space processing.
One approach to remove holes by design is to switch to geometry proxies with explicit topology, i.e., use a mesh-based model.

\paragraph{Mesh-based Representations}
Mesh-based representations explicitly model the geometry of an objects based on a set of connected geometric primitives and their appearance based on texture maps.
They have been employed, for example, to learn personalized avatars from multi-view imagery based on dynamic texture maps \cite{lombardi2018DAM}.
Differentiable rasterization approaches \cite{chen2019nvdr,loper2014opendr,kato2018neurmeshrender,liu2019softras,valentin2019tfGraphics,ravi2020pytorch3d,petersen2019pix2vex,Laine2020diffrast} enable the end-to-end integration of deep neural networks with this classical computer graphics representation.
Recently, of-the-shelf tools for differentiable rendering have been developed, e.g., TensorFlow3D \cite{valentin2019tfGraphics}, Pytorch3D \cite{ravi2020pytorch3d}, and Nvidia's nvdiffrast \cite{Laine2020diffrast}.
Differentiable rendering strategies have for example been employed for learning 3D face models \cite{tewari2017self3dmm,genova2018unsup3dmm,tran2019nonlinear3dmm} from 2D photo and video collections.
There are also techniques that store a feature map in the texture map and employ a screen-space network to compute the final image \cite{thies2019neuraltex}.
If accurate surface geometry can be obtained a-priori, mesh-based approaches are able to produce impressive results, but they often struggle if the object can not be well reconstructed.
Unfortunately, accurate 3D reconstruction is notoriously hard to acquire for humans, especially for hair, eyes, and the mouth interior.
Since such approaches require a template with fixed topology they also struggle to model topological changes and it is challenging to model occlusions in a differentiable manner.

\paragraph{Multi-Layer Representations}
One example of a mixture-based representation are Multi-Plane Images (MPIs) \cite{zhou2018mpi, tucker2020mpi, srinivasan2019mpi}.
MPIs employ a set of depth-aligned textured planes to store color and alpha information.
Novel views are synthesized by rendering the planes in back-to-front order using hardware-supported alpha blending.
These approaches are normally limited to a small restricted view-box, since specular surfaces have to be modeled via the alpha blending operator.
Local Light Field Fusion (LLFF)~\cite{mildenhall2019llff} enlarges the view-box by maintaining and smartly blending multiple local MPI-based reconstructions.
Multi-sphere images (MSIs) \cite{attal2020msi,broxton2020lightfield} replace the planar and textured geometry proxies with a set of textured concentric sphere proxies.
This enables $360^\circ$ inside-out view synthesis for VR video applications.
MatryODShka \cite{attal2020msi} enables real-time 6DoF video view synthesis for VR by converting omnidirectional stereo images to MSIs.

\paragraph{Grid-based Representations}
Grid-based representations are similar to the multi-layer representation, but are based on a dense uniform grid of voxels.
They have been extensively used to model the 3D shape of objects \cite{Mescheder2019occnet,peng2020conv,choy20163dr2n2,tulsiani2017multi,wu2016learningvox,kar2017lmvsm}.
Grid-based representations have also been used as the basis for neural rendering techniques to model object appearance \cite{sitzmann2019deepvoxels,lombardi2019nv}.
DeepVoxels~\cite{sitzmann2019deepvoxels} learns a persistent 3D feature volume for view synthesis and employs learned ray marching.
Neural Volumes \cite{lombardi2019nv} is an approach for learning dynamic volumetric reconstructions from multi-view data.
One big advantage of such representations is that they do not have to be initialized based on a fixed template mesh and are easy to optimize with gradient-based optimization techniques.
The main limiting factor for all grid-based techniques is the required cubic memory footprint.
The sparser the scene, the more voxels are actually empty, which wastes model capacity and limits resolution.
Neural Volumes employs a warping field to maximize occupancy of the template volume, but empty space is still evaluated while raymarching.
We propose to model deformable objects with a set of rigidly-moving volumetric primitives.

\paragraph{MLP-based Representations}
Multi-Layer Perceptrons (MLPs) have first been employed for modeling 3D shapes based on signed distance~\cite{park2019deepsdf, jiang2020localsdf, chabra2020deeplocalshape, saito2019pifu, saito2019pifuhd} and occupancy fields~\cite{Mescheder2019occnet, genova2020localsdf, peng2020conv}.
DeepSDF~\cite{park2019deepsdf} is one of the first works that learns the 3D shape variation of an entire object category based on MLPs.
ConvOccNet~\cite{peng2020conv} enables fitting of larger scenes by combining an MLP-based scene representation with a convolutional decoder network that regresses a grid of spatially-varying conditioning codes.
Afterwards, researchers started to also model object appearance using similar scene representations.
Neural Radiance Fields (NeRF) \cite{mildenhall2020nerf} proposes a volumetric scene representaiton based on MLPs.
One of its challenges is that the MLP has to be evaluated at a large number of sample points along each camera ray.
This makes rendering a full image with NeRF extremely slow.
Furthermore, NeRF can not model dynamic scenes or objects.
Scene Representation Networks (SRNs) \cite{sitzmann2019srn} can be evaluated more quickly as they model space with a signed-distance field. 
However, using a surface representation means that it can not represent thin structures or transparency well.
Neural Sparse Voxel Fields (NSVF)~\cite{liu2020nsvf} culls empty space based on an Octree acceleration structure, but it is extremely difficult to extend to dynamic scenes.
There exists also a large number of not-yet peer-reviewed, but impressive extensions of NeRF \cite{gafnie2929dnrf, gao2020pNeRF, Li20arxiv_nsff, martinbrualla20nerfw, park2020nerfies, rebain3020derf, tretschk2020nonrigid, Xian20arxiv_stnif, du20arxiv_nerflow, schwarz2020graf}.
While the results of MLP-based models are often visually pleasing, their main drawbacks are limited or no ability to be driven as well as their high computation cost for evaluation.

\paragraph{Summary}
Our approach is a hybrid that finds the best trade-off between volumetric- and primitive-based neural scene representations.
Thus, it produces high-quality results with fine-scale detail, is fast to render, drivable, and reduces memory constraints.

\section{Method}
\label{sec:method}
Our approach is based on a novel volumetric representation for dynamic scenes that combines the advantages of volumetric and primitive-based approaches to achieve high performance decoding and efficient rendering.
In the following, we describe our scene representation and how it can be trained end-to-end based on 2D multi-view image observations.

\subsection{Neural Scene Representation}
Our neural scene representation is inspired by primitive-based methods, such as triangular meshes, that can efficiently render high resolution models of 3D space by focusing representation capacity on occupied regions of space and ignoring those that are empty.
At the core of our method is a set of minimally-overlapping and dynamically moving volumetric primitives that together parameterize the color and opacity distribution in space over time.
Each primitive models a local region of space based on a uniform voxel grid.
This provides two main advantages that together lead to a scene representation that is highly efficient in terms of memory consumption and is fast to render:
1) fast sampling within each primitive owing to its uniform grid structure, and 
2) conditional sampling during ray marching to avoid empty space and fully occluded regions.
The primitives are linked to an underlying coarse \emph{guide} mesh (see next section) through soft constraints, but can deviate away from the mesh if this leads to improved reconstruction.
Both, the primitives' motion as well as their color and opacity distribution are parameterized by a convolutional network that enables the sharing of computation amongst them, leading to highly efficient decoding.

\subsubsection{Guide Mesh}
We employ a coarse estimate of the scene's geometry $\{\mathcal{M}_i\}_{i=1}^{N_\text{frames}}$ of every frame as basis for our scene representation.
For static scenes, it can be obtained via off-the-shelf reconstruction packages such as COLMAP~\cite{schoenberger2016sfm,schoenberger2016mvs}.
For dynamic scenes, we employ multi-view non-rigid tracking to obtain a temporally corresponded estimate of the scenes geometry over time~\cite{wu2018track}.
These meshes guide the initialization of our volumetric primitives, regularize the results, and avoid the optimization terminating in a poor local minimum.
Our model generates both the guide mesh as well as its weakly attached volumetric primitives, enabling the direct supervision of large motion using results from explicit tracking.
This is in contrast to existing volumetric methods that parameterize explicit motion via an inverse warp~\cite{lombardi2019nv,park2020nerfies}, where supervision is more challenging to employ.

\subsubsection{Mixture of Volumetric Primitives}
\label{MethodMVP}
The purpose of each of the $N_\text{prim}$ volumetric primitives is to model a small local region of 3D space.
Each volumetric primitive
$\mathcal{V}_k = \{\mathbf{t}_k, \mathbf{R}_k, \mathbf{s}_k, \mathbf{V}_k\}$ is defined by a position $\mathbf{t}_k \in \mathbb{R}^3$ in 3D space, an orientation given by a rotation matrix $\mathbf{R}_k \in \text{SO}(3)$
(computed from an axis-angle parameterization), and per-axis scale factors $\mathbf{s}_k \in \mathbb{R}^3$.
Together, these parameters uniquely describe the model-to-world transformation of each individual primitive.
In addition, each primitive contains a payload that describes the appearance and geometry of the associated region in space.
The payload is defined by a dense voxel grid $\mathbf{V}_k \in \mathbb{R}^{4 \times M_x\times M_y \times M_z}$ that stores the color (3 channels) and opacity (1 channel) for the $M_x\times M_y \times M_z$ voxels, with $M_*$ being the number of voxels along each spatial dimension. Below, we will assume our volumes are cubes with $M_*=M$ unless stated otherwise. 

As mentioned earlier, the volumetric primitives are weakly constrained to the surface of the guide mesh and are allowed to deviate from it if that improves reconstruction quality.
Specifically, their position $\mathbf{t}_k = \mathbf{\hat t}_k + \boldsymbol{\delta}_{\mathbf{t}_k}$, rotation $\mathbf{R}_k = \boldsymbol{\delta}_{\mathbf{R}_k} \cdot \mathbf{\hat R}_k$, and scale $\mathbf{s}_k = \mathbf{\hat s}_k + \boldsymbol\delta_{\mathbf{s}_k}$ are modeled relative to the guide mesh base transformation ($\mathbf{\hat t}_k$, $\mathbf{\hat R}_k$, $\mathbf{\hat s}_k$) using the regressed values ($\boldsymbol{\delta}_{\mathbf{t}_k}$, $\boldsymbol{\delta}_{\mathbf{R}_k}$, $\boldsymbol{\delta}_{\mathbf{s}_k}$).
To compute the mesh-based initialization, we generate a 2D grid in the mesh's texture space and generate the primitives at the 3D locations on the mesh that correspond to the $uv$-coordinates of the grid points.
The orientation of the primitives is initialized based on the local tangent frame of the 3D surface point they are attached to.
The scale of the boxes is initialized based on the local gradient of the $uv$-coordinates at the corresponding grid point position.
Thus, the primitives are initialized with a scale in proportion to distances to their neighbours.

\subsubsection{Opacity Fade Factor}
\label{sec:opacity_fade_factor}
Allowing the volumetric primitives to deviate from the guide mesh is important to account for deficiencies in the initialization strategy, low guide mesh quality, and insufficient coverage of objects in the scene.  
However, allowing for motion is not enough; during training the model can only receive gradients from regions of space that the primitives cover, resulting in a limited ability to self assemble and expand to attain more coverage of the scene's content.
Furthermore, it is easier for the model to reduce opacity in empty regions than to move the primitives away. This wastes primitives that would be better-utilized in regions with higher occupancy.
To mitigate this behavior, we apply a windowing function $\mathbf{W} \in \mathbb{R}^{M^3}$ to the opacity of the payload that takes the form:
\begin{equation}
    \mathbf{W}(x,y,z) = \exp{\left\{-\alpha \left ( x^{\beta} + y^{\beta} + z^{\beta} \right ) \right \}} \enspace,
\end{equation}
where $(x,y,z) \in [-1,1]^3$ are normalized coordinates within the primitive's payload volume.
Here, $\alpha$ and $\beta$ are hyperparameters that control the rate of opacity decay towards the edges of the volume.
This windowing function adds an inductive bias to explain the scene's contents via motion instead of payload since the magnitude of gradients that are propagated through opacity values at the edges of the payload are downscaled.
We note that this does not prevent the edges of the opacity payload from being able to take on large values, rather, our construction forces them to learn more slowly~\cite{Karras2019stylegan2}, thus favoring motion of the primitives whose gradients are not similarly impeded.
We found $\alpha=8$ and $\beta=8$ was a good balance between scene coverage and reconstruction accuracy and keep them fixed for all experiments.

\subsubsection{Network Architecture}
\begin{figure*}[t]
    \centering
    \includegraphics[width=\linewidth]{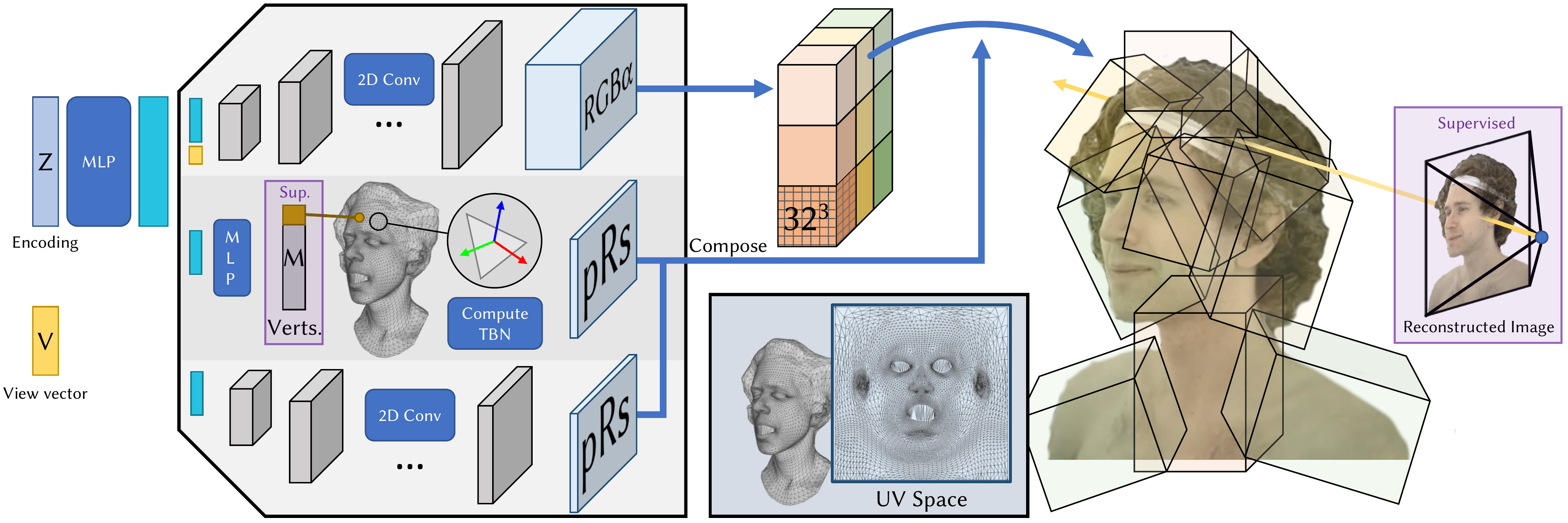}
    \vspace{-0.6cm}
    \caption{
    Decoder architecture.
    We decode latent code $\mathbf{z}$ into 1) a \emph{slab} of volumetric primitives, 2) vertex positions of a guide mesh, which is used to compute a base transformation for each primitive ($\mathbf{\hat t}_k$, $\mathbf{\hat R}_k$, $\mathbf{\hat s}_k$) by computing a rotation matrix from the tangent, bitangent, normal vectors at points evenly distributed in the UV space of the mesh, and 3) a residual transformation ($\boldsymbol{\delta}_{\mathbf{t}_k}$, $\boldsymbol{\delta}_{\mathbf{R}_k}$, $\boldsymbol{\delta}_{\mathbf{s}_k}$).
    The 3D slab is divided into individual primitives, and transformed into world space by composing the two transformations.
    We raymarch through the volumetric primitives, accumulating color and opacity, to form the reconstructed image.
    Both the reconstructed image and the mesh vertex positions are supervised during training from ground truth data.
    }
    \label{fig:DecoderArchitecture}
\end{figure*}

We employ an encoder-decoder network to parameterize the coarse tracked proxy mesh as well as the weakly linked mixture of volumetric primitives.
Our approach is based on Variational Autoencoders (VAEs) \cite{kingma2013vae} to encode the dynamics of the scene using a low-dimensional latent code $\mathbf{z}\in\mathbb{R}^{256}$.
Note that the goal of our construction is only to produce the decoder. The role of the encoder during training is to encourage a well structured latent space. It can be discarded upon training completion and replaced with an application specific encoder \cite{wei2019VR} or simply with latent space traversal \cite{Abdal_2019_ICCV}. 
In the following, we provide details of the encoder for different training settings as well as our four decoder modules.

\paragraph{Encoder}
The architecture of our encoder is specialized to the data available for modeling. When coarse mesh tracking is available, we follow the architecture in~\cite{lombardi2018DAM}, which takes as input the tracked geometry and view-averaged unwarped texture for each frame. Geometry is passed through a fully connected layer, and texture through a convolutional branch, before being fused and further processed to predict the parameters of a normal distribution $\mathcal{N(\boldsymbol\mu, \boldsymbol\sigma)}$, where $\boldsymbol\mu \in \mathbb{R}^{256}$ is the mean and $\boldsymbol\sigma \in \mathbb{R}^{256}$ is the standard deviation.
When tracking is not available we follow the encoder architecture in~\cite{lombardi2019nv}, where images from $K=1$ fixed view is taken as input.
To learn a smooth latent space with good interpolation properties, we regularize using a KL-divergence loss that forces the predicted distribution to stay close to a standard normal distribution.
The latent vector $\mathbf{z}\in\mathbb{R}^{256}$ is obtained by sampling from the predicted distribution using the reparameterization trick \cite{kingma2013vae}.

\paragraph{Decoder}
Our decoder is comprised of four modules; two geometry decoders and two payload decoders. The geometry decoders determine the primitives' model-to-world transformations. 
$
\mathcal{D_{\text{mesh}}}:
\mathbb{R}^{256}
\rightarrow
\mathbb{R}^{3\times N_{\text{mesh}}}
$
predicts the guide mesh $\mathcal{M}$ used to initialize the transformations. It is comprised of a sequence of fully connected layers. 
$
\mathcal{D_{\text{pRs}}}:
\mathbb{R}^{256}
\rightarrow
\mathbb{R}^{9 \times N_{\text{prim}}}
$
is responsible for predicting the deviations in position, rotation (as a Rodrigues vector), and scale ($\boldsymbol{\delta}_{\mathbf{t}_k}$, $\boldsymbol{\delta}_{\mathbf{R}_k}$, $\boldsymbol{\delta}_{\mathbf{s}_k}$) from the guide mesh initialization. It uses a 2D convolutional architecture to produces the motion parameters as channels of a 2D grid following the primitive's ordering in the texture's uv-space described in \S\ref{MethodMVP}. 
The payload decoders determine the color and opacity stored in the primitives' voxel grid $\mathbf{V}_k$. 
$
\mathcal{D_{\alpha}}:
\mathbb{R}^{256}
\rightarrow
\mathbb{R}^{1 \times M^3 \times N_{\text{prim}}}
$
computes opacity based on a 2D convolutional architecture. 
$
\mathcal{D_{\text{rgb}}}:
\mathbb{R}^{256+3}
\rightarrow
\mathbb{R}^{3 \times M^3 \times N_{\text{prim}}}
$
computes view-dependent RGB color. It is also based on 2D transposed convolutions and uses an object-centric view-vector $\mathbf{d}_{\text{view}} \in \mathbb{R}^3$, i.e., a vector pointing to the center of the object/scene.
The view vector allows the decoder to model view-dependent phenomena, like specularities.
Unlike the geometry decoders, which employ small networks and are efficient to compute, payload decoders present a significant computational challenge due to the total number of elements they have to generate. Our architecture, shown in Fig.~\ref{fig:DecoderArchitecture}, addresses this by avoiding redundant computation through the use of a convolutional architecture. Nearby locations in the output \emph{slab} of $\mathbf{V}_k$'s leverage shared features from earlier layers of the network. This is in contrast to MLP-based methods, such as~\cite{mildenhall2020nerf}, where each position requires independent computation of all features in all layers, without any sharing. Since our texture space is the result of a mesh-atlasing algorithm that tends to preserve the locality structure of the underlying 3D mesh, the regular grid ordering of our payload $\mathbf{V}_k$ within the decoded slab (see \S\ref{MethodMVP}) well leverages the spatially coherent structures afforded by devonvolution. The result is an efficient architecture with good reconstruction capacity.

\paragraph{Background Model}
MVP is designed to model objects in a scene from an outside-in camera configuration, but the extent of object coverage is not know a-priori.
Thus, we need a mechanism for separating objects from the backgrounds in the scene.
However, existing segmentation algorithms can fail to capture fine details around object borders and can be inconsistent in 3D.
Instead, we jointly model the objects as well as the scene's background.
Whereas the objects are modeled using MVP, we use a separate neural network to model the background as a modulation of images captured of the empty scene with the objects absent.
Specifically, our background model for the $i^{\text{th}}$-camera takes the form:
\begin{equation}
    \mathcal{B}_i(\mathbf{x}) = \text{softplus}\left\{ \mathcal{\bar{B}}_i + F_{\theta}(\mathbf{c}_i, \mathbf{d}_i(\mathbf{x})) \right \} \enspace,
\end{equation}
where $\mathcal{\bar{B}}_i$ is the image of the empty capture space, $\mathbf{c}_i$ is the camera center and $\mathbf{d}_i$ is the ray direction for pixel $\mathbf{x}$.
The function $F$ is an MLP with weights $\theta$ that takes position-encoded camera coordinates and ray directions and produces a rgb-color using an architecture similar to NeRF~\cite{mildenhall2020nerf}.
The background images of the empty scene are not sufficient by themselves since objects in the scene can have effects on the background, which, if not accounted for, are absorbed in to the MVP resulting in hazy reconstructions as observed in NV~\cite{lombardi2019nv}.
Examples of these effects include shadowing and content outside of the modeling volume, like supporting stands and chairs.
As we will see in \S\ref{DifferentiableVolumetricAggregation}, MVP rendering produces an image with color, $\mathcal{I}$, and alpha, $\mathcal{A}$, channels.
These are combined with the background image to produce the final output that is compared to the captured images during training through alpha-compositing:
$\mathcal{\tilde{I}}_i = \mathcal{A}_i \ \mathcal{I}_i + (1-\mathcal{A}_i) \ \mathcal{B}_i$.

\subsection{Efficient and Differentiable Image Formation}
The proposed scene representation is able to focus the representational power of the encoder-decoder network on the occupied regions of 3D space, thus leading to a high resolution model and efficient decoding.
However, we still need to be able to efficiently render images using this representation. 
For this, we propose an approach that combines an efficient raymarching strategy with a differentiable volumetric aggregation scheme.

\subsubsection{Efficient Raymarching}
To enable efficient rendering, our algorithm should:
1) skip samples in empty space, and
2) employ efficient payload sampling.
Similar to~\cite{lombardi2019nv}, the regular grid structure of our payload enables efficient sampling via trilinear interpolation. However, in each step of the ray marching algorithm, we additionally need to find within which primitives the current evaluation point lies. These primitives tend to be highly irregular with positions, rotations, and scales that vary on a per-frame basis. For this, we employ a highly optimized data-parallel BVH implementation \cite{karras2013bvh} that requires less than $0.1$ ms for 4096 primitives at construction time.
This enables us to rebuild the BVH on a per-frame basis, thus handling dynamic scenes, and provides us with efficient intersection tests.
Given this data structure of the scene, we propose a strategy for \emph{limiting} evaluations as much as possible.
First, we compute and store the primitives that each ray intersects. 
We use this to compute $(t_\text{min}$, $t_\text{max}$), the domain of integration.
While marching along a ray between $t_\text{min}$ and $t_\text{max}$, we check each sample only against the ray-specific list of intersected primitives. 
Compared to MLP-based methods, e.g., NeRF \cite{mildenhall2020nerf}, our approach exhibits very fast sampling. 
If the number of overlapping primitives is kept low, the total sampling cost is much smaller than a deep MLP evaluation at each step, which is far from real-time even with a good importance sampling strategy.

\subsubsection{Differentiable Volumetric Aggregation}
\label{DifferentiableVolumetricAggregation}
We require a differentiable image formation model to enable end-to-end training based on multi-view imagery.
Given the sample points in occupied space extracted by the efficient ray marching strategy, we employ an accumulative volume rendering scheme as in~\cite{lombardi2019nv} that is motivated by front-to-back additive alpha blending.
During this process, the ray accumulates color as well as opacity.
Given a ray $r_\mathbf{p}(t) =\mathbf{o}_\mathbf{p} + t \mathbf{d}_\mathbf{p}$ with starting position $\mathbf{o}_\mathbf{p}$ and ray direction $\mathbf{d}_\mathbf{p}$, we solve the following integral using numerical quadrature:
$$
\mathcal{I}_p = 
\int_{t_\text{min}}^{t_\text{max}}{
\mathbf{V}_{\text{col}}(r_\mathbf{p}(t)) \cdot \frac{dT(t)}{dt}\cdot dt}
\enspace .
$$
$$
T(t)=
\min\Big(
\int_{t_\text{min}}^{t}{
\mathbf{V}_{\alpha}(r_\mathbf{p}(t))\cdot dt
, 1\Big)}
\enspace .
$$
Here, $\mathbf{V}_{\text{col}}$ and $\mathbf{V}_{\alpha}$ are the global color and opacity field computed based on the current instantiation of the volumetric primitives.
We set the alpha value associated with the pixel to  $\mathcal{A}_p = T(t_{\text{max}})$.
For high performance rendering, we employ an early stopping strategy based on the accumulated opacity, i.e., if the accumulated opacity is larger than  $1.0-\epsilon_{\text{early}}$ we terminate ray marching, since the rest of the sample points do not have a significant impact on the final pixel color.
If a sample point is contained within multiple volumetric primitives, we combine their values in their BVH order based on the accumulation scheme.
Our use of the additive formulation for integration, as opposed to the multiplicative form~\cite{mildenhall2020nerf}, is motivated by its independence to ordering up to the saturation point.  
This allows for a backward pass implementation that is more memory efficient, since we do not need to keep the full $O(n^2)$ graph of operations.
Thus, our implementation requires less memory and allows for larger batch sizes during training.
For more details, we refer to the supplemental document.

\subsection{End-to-end Training}
\label{sec:end_to_end_training}
Next, we discuss how we can train our approach end-to-end based on a set of 2D multi-view images.
The trainable parameters of our model are $\Theta$.
Given a multi-view video sequence $\{\mathcal{I}^{(i)}\}_{i=1}^{N_{img}}$ with $N_{\text{img}}=N_{\text{frames}} \cdot N_{\text{cams}}$ training images, our goal is to find the optimal parameters $\Theta^*$ that best explain the training data.
To this end, we solve the following optimization problem:
$$
\Theta^* = \argmin_{\Theta}
{\sum_{i=0}^{N_{\text{img}}-1}
{\sum_{p=0}^{N_{\text{pixels}}-1}
{\mathcal{L}\big(\Theta;\mathcal{I}^{(i)}_{p}\big)}}}
\enspace .
$$
We employ ADAM \cite{KingmaB14} to solve this optimization problem based on stochastic mini-batch optimization.
In each iteration, our training strategy uniformly samples rays from each image in the current batch to define the loss function.
We employ a learning rate $lr=0.0001$ and all other parameters are set to their default values.
Our training objective is of the following from:
$$
\mathcal{L}(\Theta;\mathcal{I}_{p}) =
\mathcal{L}_{\text{pho}}(\Theta;\mathcal{I}_{p}) + \mathcal{L}_{\text{geo}}(\Theta) + \mathcal{L}_\text{vol}(\Theta) + \mathcal{L}_\text{del}(\Theta) + \mathcal{L}_\text{kld}(\Theta)
\enspace .
$$
It consists of a photometric reconstruction loss $\mathcal{L}_{\text{pho}}$, a coarse geometry reconstruction loss $\mathcal{L}_{\text{geo}}$, a volume minimization prior $\mathcal{L}_\text{vol}$, a delta magnitude prior $\mathcal{L}_\text{del}$, and a Kullback–Leibler (KL) divergence prior $\mathcal{L}_\text{kld}$ to regularize the latent space of our Variational Autoencoder (VAE) \cite{kingma2013vae}.
In the following, we provide more details on the individual energy terms.

\begin{figure*}[t]
    \centering
    \includegraphics[width=\linewidth]{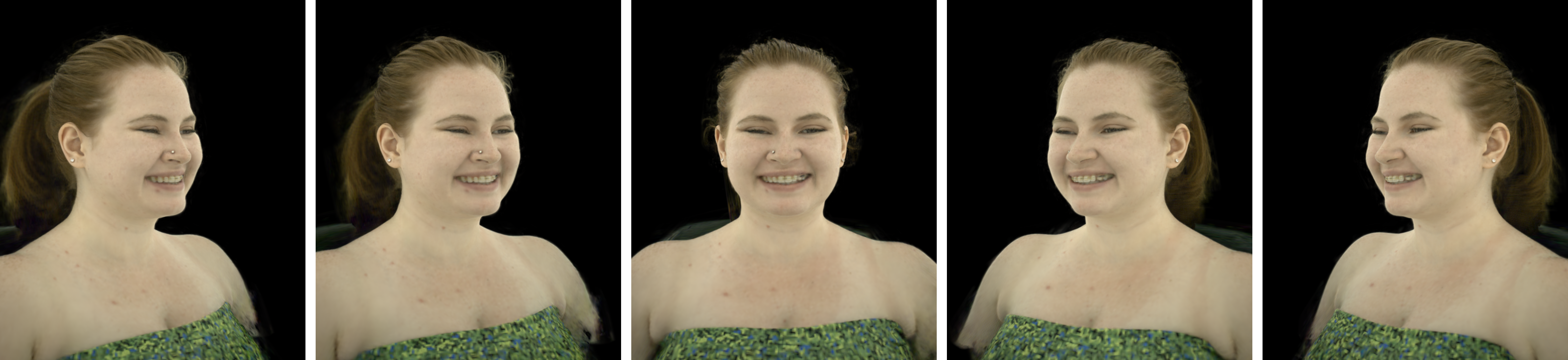}
    \vspace{-0.8cm}
    \caption{Novel view synthesis example. Our novel approach enables high fidelity novel view synthesis of a complete, dynamic, upper body at real-time rates.}
    \label{fig:results:nvs}
\end{figure*}

\begin{figure*}[t]
    \centering
    \includegraphics[width=\linewidth]{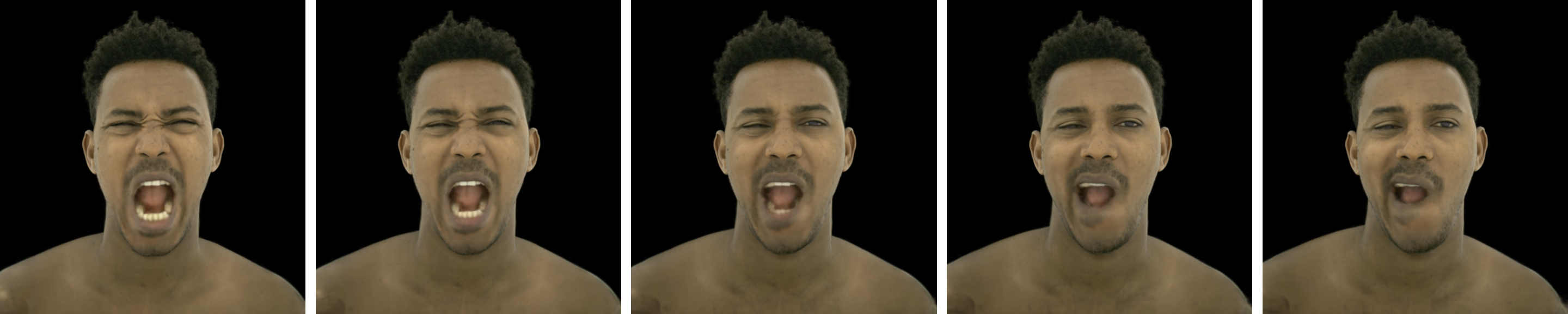}
    \vspace{-0.8cm}
    \caption{
    Latent space interpolation example. Our approach enables animating the reconstructions via latent-space interpolation between two keyframes.
    Even for extremely challenging facial expressions the interpolated results are highly realistic.
    The output of our motion model, which produces rotations, translations, and scales, is effectively a forward warp. Such a warp can be sensibly-interpolated whereas inverse warps cannot.
    }
    \label{fig:results:interp}
\end{figure*}

\paragraph{Photometric Reconstruction Loss}
We want to enforce that the synthesized images look photo-realistic and match the ground truth.
To this end, we compare the synthesized pixels $\mathcal{\bar I}_p(\Theta)$ to the ground truth $\mathcal{I}_p$ using the following loss function:
$$
\mathcal{L}_{\text{pho}} = \lambda_{\text{pho}}
\frac{1}{N_\mathcal{P}} \sum_{p \in \mathcal{P}}{
\big|\big| \mathcal{I}_p - \mathcal{\bar I}_p(\Theta) \big|\big|_2^2}
\enspace .
$$
Here, $\mathcal{P}$ is the set of sampled pixels and $w_p$ is a per-pixel weight.
We set a relative weight of $\lambda_{\text{photo}}=1.0$.

\paragraph{Mesh Reconstruction Loss}
We also want to enforce that the coarse mesh proxy follows the motion in the scene.
To this end, we compare the regressed vertex positions to the available ground truth traced mesh using the following loss function:
$$
\mathcal{L}_{\text{geo}} = \lambda_{\text{geo}} \frac{1}{N_{\text{mesh}}}
\sum_{i=0}^{N_{\text{mesh}}}{\big|\big| \mathbf{v}_i - \mathbf{\bar v}_i(\Theta) \big|\big|_2^2} \enspace .
$$
Here, we employ an $\ell_2$-loss function, $\mathbf{v}_i$ is the ground truth position of the tracked mesh, and $\mathbf{\bar v}_i(\Theta)$ is the corresponding regressed vertex position.
We employ the coarse mesh-based tracking used in the approach of \cite{lombardi2018DAM}.
The mesh reconstruction loss pulls the volumetric primitives, which are weakly linked to it, to an approximately correct position.
Note, the primitives are only weakly linked to the mesh proxy and can deviate from their initial positions if that improves the photometric reconstruction loss.
We set a relative weight of $\lambda_{\text{geo}}=0.1$.

\paragraph{Volume Minimization Prior}
We constrain the volumetric primitives to be as small as possible.
The reasons for this are twofold:
1) We want to prevent them from overlapping too much, since this wastes model capacity in already well explained regions, and
2) We want to prevent loss of resolution by large primitives overlapping empty space.
To this end, we employ the following volume minimization prior:
$$
\mathcal{L}_{\text{vol}}=\lambda_{\text{vol}}
\sum_{i=1}^{N_\text{prim}}{\text{Prod}( \mathbf{s}_i)} \enspace .
$$
Here, $\mathbf{s}_i = \mathbf{\hat s}_i + \boldsymbol{\delta}_{\mathbf{s}_i}$ is the vector of side lengths of the primitive and $\text{Prod}(\bullet)$ is the product of the values of a vector, e.g., in our case the volume of a primitive.
We minimize the total volume with a relative weight of $\lambda_{\text{vol}}=0.01$.

\section{Results}
In this section we describe our datasets for training and evaluation, present results on several challenging sequences, perform ablation studies over our model's components, and compare to the state of the art. 
We perform both qualitative and quantitative evaluations.

\subsection{Training Data}
We evaluate our approach on a large number of sequences captured using a spherically arranged multi-camera capture system with $\approx 100$ synchronized color cameras.
Each dataset contains roughly 25,000 frames from each camera.
The cameras record with a resolution of $2668 \times 4096$ at $30$Hz, and are equally distributed on the surface of the spherical structure with a radius of $1.2$ meters.
They are geometrically calibrated with respect to each other with the intrinsic and extrinsic parameters of a pinhole camera model.
For training and evaluation, we downsample the images to a resolution of $667\times 1024$ to reduce the time it takes to load images from disk, 
keeping images from all but 8 cameras for training, with the remainder used for testing. 
To handle the different radiometric properties of the cameras, e.g.,~color response and white balance, we employ per-camera color calibration based on $6$ parameters (gain and bias per color channel)
similar to~\cite{lombardi2018DAM}, but pre-trained for all cameras once for each dataset.
We train each scene for 500,000 iterations, which takes roughly 5 days on a single NVIDIA Tesla V100.

\subsection{Qualitative Results}
Our approach achieves a high level of fidelity while matching the completeness of volumetric representations, e.g., hair coverage and inner mouth, see Fig.~\ref{fig:results:nvs}, but with an efficiency closer to mesh-based approaches.
Our fastest model is able to render binocular stereo views at a resolution of $896 \times 832$ at $40$Hz, which enables live visualization of our results in a virtual reality setting; please see the supplemental video for these results.
Furthermore, our model can represent dynamic scenes, supporting free view-point video applications, see Fig.~\ref{fig:results:nvs}.
Our model also enables animation, which we demonstrate via latent-space interpolation, see Fig.~\ref{fig:results:interp}.
Due to the combination of the variational architecture with a forward warp, our approach produces highly realistic animation results even if the facial expressions in the keyframes are extremely different.
\begin{figure*}[h]
    \centering
    \includegraphics[width=\linewidth]{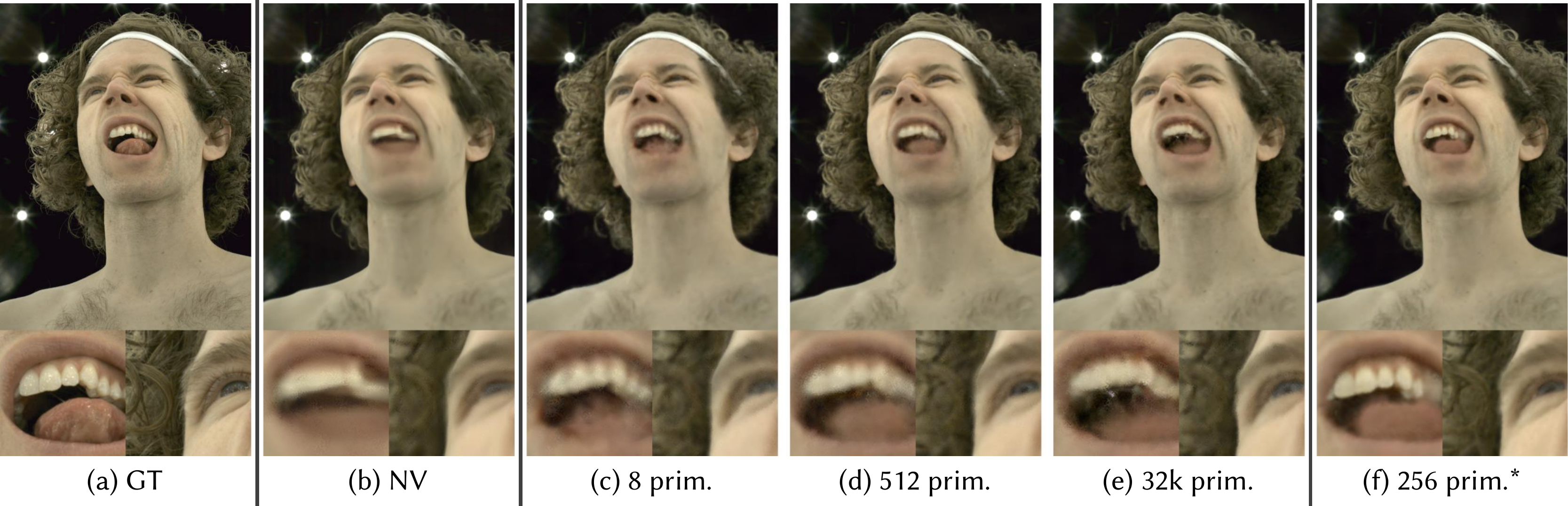}
    \vspace{-0.8cm}
    \caption{
    We evaluate results by comparing models with a varying number of primitives on held out views while keeping the total number of voxels constant (2 million).
    From left to right: ground truth, neural volumes, 8 primitives, 512 primitives, and 32768 primitives.
    We find that hundreds of primitives gives the best balance of quality and performance, with higher numbers of primitives exhibiting fine detail, but struggling with elaborate motion, e.g.,~mouth interior.
    The last column shows our final model, which uses 256 primitives and 8 million total voxels. (*): only 2D transposed convolutions.
    }
    \label{fig:results:boxsize}
\end{figure*}
\begin{table*}[h!]
\centering
 \begin{tabular}{r | c | c c c c c c | c c }
 & NV & 8 prim. & 64 prim. & 512 prim. & 4k prim. & 32k prim. & 262k prim. & 256 prim.* & 256 prim.** \\
 \hline
 MSE ($\downarrow$)   & \cellcolor[rgb]{0.99,0.50,0.00} 49.1535
 & \cellcolor[rgb]{1.00,0.61,0.00} 46.5067
 & \cellcolor[rgb]{1.00,0.79,0.03} 42.2567
 & \cellcolor[rgb]{1.00,0.81,0.04} 41.9601
 & \cellcolor[rgb]{1.00,0.90,0.10} 40.3211
 & \cellcolor[rgb]{1.00,0.88,0.08} 40.6524
 & \cellcolor[rgb]{1.00,0.73,0.00} 43.4307
 & \cellcolor[rgb]{0.99,0.92,0.13} 39.9031
 & \cellcolor[rgb]{0.89,1.00,0.48} \textbf{37.0805}
\\
 PSNR ($\uparrow$)   & \cellcolor[rgb]{0.99,0.50,0.00} 31.2153
 & \cellcolor[rgb]{1.00,0.60,0.00} 31.4556
 & \cellcolor[rgb]{1.00,0.77,0.02} 31.8718
 & \cellcolor[rgb]{1.00,0.78,0.02} 31.9024
 & \cellcolor[rgb]{1.00,0.88,0.08} 32.0755
 & \cellcolor[rgb]{1.00,0.86,0.07} 32.0399
 & \cellcolor[rgb]{1.00,0.71,0.00} 31.7528
 & \cellcolor[rgb]{1.00,0.90,0.10} 32.1207
 & \cellcolor[rgb]{0.89,1.00,0.48} \textbf{32.4393}
\\
 SSIM ($\uparrow$)   & \cellcolor[rgb]{0.99,0.50,0.00} 0.9293
 & \cellcolor[rgb]{0.99,0.54,0.00} 0.9301
 & \cellcolor[rgb]{1.00,0.71,0.00} 0.9336
 & \cellcolor[rgb]{1.00,0.69,0.00} 0.9333
 & \cellcolor[rgb]{1.00,0.80,0.04} 0.9352
 & \cellcolor[rgb]{1.00,0.78,0.02} 0.9349
 & \cellcolor[rgb]{1.00,0.57,0.00} 0.9308
 & \cellcolor[rgb]{1.00,0.74,0.00} 0.9344
 & \cellcolor[rgb]{0.89,1.00,0.48} \textbf{0.9393}
\\
 LPIPS ($\downarrow$)   & \cellcolor[rgb]{1.00,0.74,0.00} 0.2822
 & \cellcolor[rgb]{0.99,0.50,0.00} 0.3151
 & \cellcolor[rgb]{1.00,0.70,0.00} 0.2879
 & \cellcolor[rgb]{1.00,0.80,0.03} 0.2764
 & \cellcolor[rgb]{1.00,0.81,0.04} 0.2755
 & \cellcolor[rgb]{1.00,0.89,0.09} 0.2670
 & \cellcolor[rgb]{1.00,0.79,0.03} 0.2767
 & \cellcolor[rgb]{1.00,0.67,0.00} 0.2921
 & \cellcolor[rgb]{0.89,1.00,0.48} \textbf{0.2484}
\\
\hline
 decode ($\downarrow$)   & \cellcolor[rgb]{0.99,0.53,0.00} 54.8993
 & \cellcolor[rgb]{0.99,0.53,0.00} 55.3351
 & \cellcolor[rgb]{0.99,0.50,0.00} 57.5364
 & \cellcolor[rgb]{0.99,0.54,0.00} 54.0634
 & \cellcolor[rgb]{1.00,0.72,0.00} 39.3384
 & \cellcolor[rgb]{1.00,0.72,0.00} 39.6899
 & \cellcolor[rgb]{0.98,0.93,0.17} 26.4612
 & \cellcolor[rgb]{0.92,0.98,0.40} 20.5428
 & \cellcolor[rgb]{0.89,1.00,0.48} \textbf{18.3311}
\\
 raymarch ($\downarrow$)   & \cellcolor[rgb]{0.89,1.00,0.48} \textbf{7.0539}
 & \cellcolor[rgb]{0.90,1.00,0.47} 7.3450
 & \cellcolor[rgb]{0.92,0.98,0.40} 8.7660
 & \cellcolor[rgb]{0.94,0.96,0.30} 10.6198
 & \cellcolor[rgb]{0.98,0.92,0.16} 13.3397
 & \cellcolor[rgb]{1.00,0.78,0.03} 19.8951
 & \cellcolor[rgb]{0.99,0.51,0.00} 35.8098
 & \cellcolor[rgb]{0.95,0.95,0.27} 11.0970
 & \cellcolor[rgb]{0.99,0.50,0.00} 36.6212
\\
 total ($\downarrow$)   & \cellcolor[rgb]{0.99,0.56,0.00} 61.9532
 & \cellcolor[rgb]{0.99,0.55,0.00} 62.6801
 & \cellcolor[rgb]{0.99,0.50,0.00} 66.3024
 & \cellcolor[rgb]{0.99,0.52,0.00} 64.6831
 & \cellcolor[rgb]{1.00,0.69,0.00} 52.6781
 & \cellcolor[rgb]{1.00,0.60,0.00} 59.5850
 & \cellcolor[rgb]{0.99,0.56,0.00} 62.2710
 & \cellcolor[rgb]{0.89,1.00,0.48} \textbf{31.6398}
 & \cellcolor[rgb]{1.00,0.66,0.00} 54.9523
\\
\end{tabular}
 \caption{
 Quantitative evaluation of number of volumetric primitives. We evaluate quality (MSE, PSNR, SSIM, LPIPS) and execution time (decode time, raymarch time, and total time in milliseconds) on held out views. Quality generally increases as more primitives are used, as they can more tightly fit the geometry of the surface. The performance drops for extreme numbers of primitives, since this increases the difficulty of the learning problem. Raymarching becomes more costly as more primitives are used, as the overhead of checking if a sample point lies in each primitive dominates the sampling time. Decode time decreases with number of primitives (note the number of voxels remains constant at 2 million) as the ``slab'' has less thickness and therefore requires fewer 3D convolutions, which tend to be more expensive. The rightmost two columns show our final optimized model, which uses 256 primitives, 8 million voxels, and 2D convolutions exclusively. (*): only 2D convolutions, and (**): smaller raymarching step size.
 }
 \label{tab:boxes}
 \vspace{-2mm}
\end{table*}

\begin{figure}[t]
    \centering
    \includegraphics[width=\linewidth]{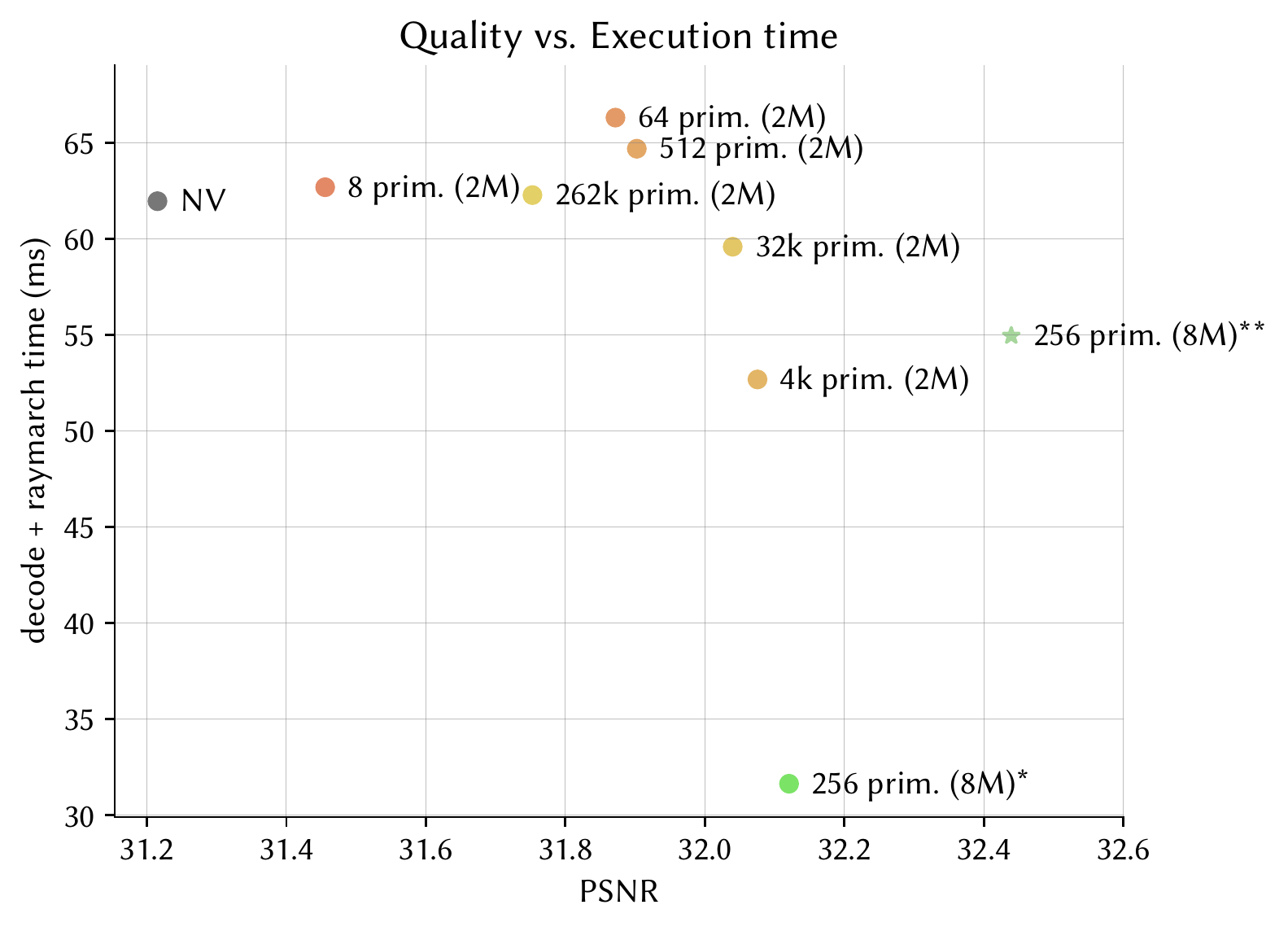}
    \vspace{-0.8cm}
    \caption{Quality vs.~Execution time for varying number of primitives and model architectures. We show our base model (8-32k primitives with 2 million total voxels) and our optimized model (256 primitives with 8 million total voxels) compared to Neural Volumes. Our optimized model greatly improved performance over Neural Volumes in both quality (PSNR) and total execution time.}
    \label{fig:results:qualityvruntime}
\end{figure}

\subsection{Ablation Studies}
We perform a number of ablation studies to support each of our design choices.

\paragraph{Number of Primitives}
We investigated the influence the number of primitives, $N_{\text{prim}}$, has on rendering quality and runtime performance.
A quantitative evaluation can be found in Tab.~\ref{tab:boxes}.
Here, we compared models with varying number of primitives on held out views, while keeping the total number of voxels constant ($\sim 2$ million).
In total, we compared models with 1, 8, 64, 512, 4096, 32768, and 262144 primitives.
Note that the special case of exactly one volumetric primitive corresponds to the setting used in \citet{lombardi2019nv}.
Our best model uses 256 primitives and $8$ million voxels.
Fig.~\ref{fig:results:qualityvruntime} shows this data in a scatter plot with PSNR on the x-axis and total execution time on the y-axis.
Fig.~\ref{fig:results:boxsize} shows the qualitative results for this evaluation.
As the results show, if the number of primitives is too low results appear blurry, and if the number of primitives is too large the model struggles to model elaborate motion, e.g.,~in the mouth region, leading to artifacts.

\paragraph{Primitive Volume Prior}
\begin{figure}[t]
    \centering
    \includegraphics[width=\linewidth]{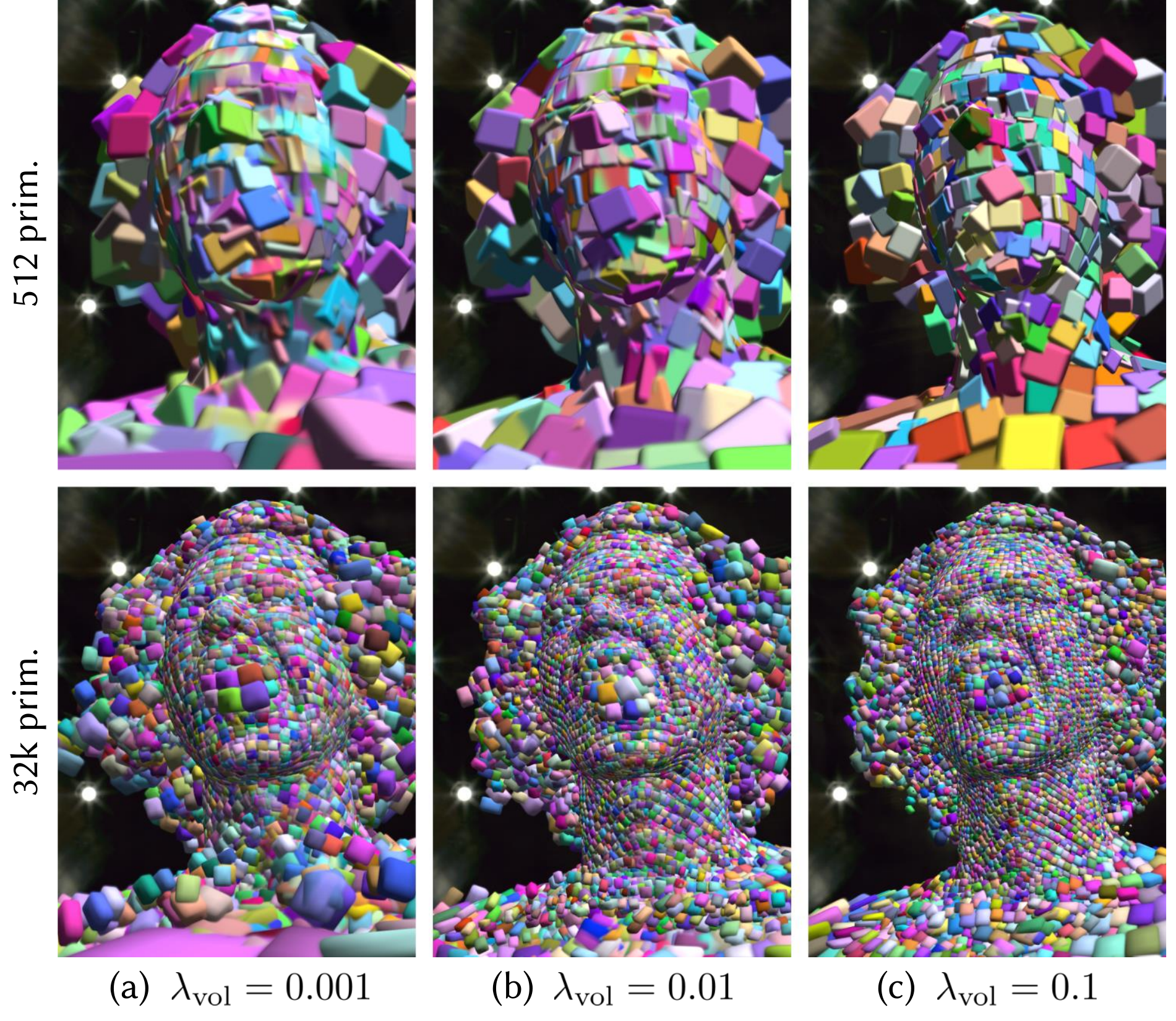} 
    \vspace{-0.8cm}
    \caption{
    We evaluate different strengths of the primitive volume prior for 512 primitives and 32768 primitives.
    A stronger primitive volume prior leads to less overlap and thus speeds up raymarching.
    If the prior is too strong, however, holes appear and the reconstruction error increases.
    }
    \label{fig:results:boxprior}
\end{figure}
\begin{table}[h]
\centering
 \begin{tabular}{r | c c c}
 & \shortstack{512 prim. \\ $\lambda_{\text{vol}}=0.001$} & \shortstack{512 prim. \\ $\lambda_{\text{vol}}=0.01$} & \shortstack{512 prim. \\ $\lambda_{\text{vol}}=0.1$} \\
 \hline
 MSE ($\downarrow$)   & \cellcolor[rgb]{0.91,0.98,0.41}  42.3944
 & \cellcolor[rgb]{0.89,1.00,0.48}  \textbf{41.9601}
 & \cellcolor[rgb]{0.99,0.50,0.00}  51.9607
\\
 PSNR ($\uparrow$)   & \cellcolor[rgb]{0.91,0.98,0.41}  31.8577
 & \cellcolor[rgb]{0.89,1.00,0.48}  \textbf{31.9024}
 & \cellcolor[rgb]{0.99,0.50,0.00}  30.9740
\\
 SSIM ($\uparrow$)   & \cellcolor[rgb]{0.89,1.00,0.48}  \textbf{0.9383}
 & \cellcolor[rgb]{1.00,0.82,0.05}  0.9333
 & \cellcolor[rgb]{0.99,0.50,0.00}  0.9253
\\
 LPIPS ($\downarrow$)   & \cellcolor[rgb]{1.00,0.68,0.00}  0.4119
 & \cellcolor[rgb]{0.89,1.00,0.48}  \textbf{0.2764}
 & \cellcolor[rgb]{0.99,0.50,0.00}  0.4928
\\
\hline
 decode ($\downarrow$)   & \cellcolor[rgb]{0.99,0.50,0.00}  55.0023
 & \cellcolor[rgb]{0.89,1.00,0.48}  \textbf{54.0634}
 & \cellcolor[rgb]{1.00,0.89,0.09}  54.3226
\\
 raymarch ($\downarrow$)   & \cellcolor[rgb]{0.99,0.50,0.00}  16.1322
 & \cellcolor[rgb]{1.00,0.86,0.07}  10.6198
 & \cellcolor[rgb]{0.89,1.00,0.48}  \textbf{7.8910}
\\
 total ($\downarrow$)   & \cellcolor[rgb]{0.99,0.50,0.00}  71.1345
 & \cellcolor[rgb]{1.00,0.89,0.09}  64.6831
 & \cellcolor[rgb]{0.89,1.00,0.48}  \textbf{62.2137}
\\
 \hline \hline
 & \shortstack{32k prim. \\ $\lambda_{\text{vol}}=0.001$} & \shortstack{32k prim. \\ $\lambda_{\text{vol}}=0.01$} & \shortstack{32k prim. \\ $\lambda_{\text{vol}}=0.1$} \\
 \hline
 MSE ($\downarrow$)   & \cellcolor[rgb]{0.96,0.94,0.24}  41.9728
 & \cellcolor[rgb]{0.89,1.00,0.48}  \textbf{40.6524}
 & \cellcolor[rgb]{0.99,0.50,0.00}  48.9140
\\
 PSNR ($\uparrow$)   & \cellcolor[rgb]{0.97,0.94,0.22}  31.9011
 & \cellcolor[rgb]{0.89,1.00,0.48}  \textbf{32.0399}
 & \cellcolor[rgb]{0.99,0.50,0.00}  31.2365
\\
 SSIM ($\uparrow$)   & \cellcolor[rgb]{0.89,1.00,0.48}  \textbf{0.9357}
 & \cellcolor[rgb]{0.92,0.98,0.38}  0.9349
 & \cellcolor[rgb]{0.99,0.50,0.00}  0.9238
\\
 LPIPS ($\downarrow$)   & \cellcolor[rgb]{1.00,0.71,0.00}  0.3810
 & \cellcolor[rgb]{0.89,1.00,0.48}  \textbf{0.2670}
 & \cellcolor[rgb]{0.99,0.50,0.00}  0.4708
\\
\hline
 decode ($\downarrow$)   & \cellcolor[rgb]{0.89,1.00,0.48}  \textbf{35.8994}
 & \cellcolor[rgb]{0.99,0.50,0.00}  39.6899
 & \cellcolor[rgb]{0.99,0.57,0.00}  39.1676
\\
 raymarch ($\downarrow$)   & \cellcolor[rgb]{0.99,0.50,0.00}  48.3455
 & \cellcolor[rgb]{1.00,0.90,0.10}  19.8951
 & \cellcolor[rgb]{0.89,1.00,0.48}  \textbf{9.6950}
\\
 total ($\downarrow$)   & \cellcolor[rgb]{0.99,0.50,0.00}  84.2449
 & \cellcolor[rgb]{1.00,0.87,0.08}  59.5850
 & \cellcolor[rgb]{0.89,1.00,0.48}  \textbf{48.8626}
\\
 \end{tabular}
 %
 \caption{
 Ablation of primitive volume prior.
 The volume prior provides a trade-off between quality and performance, with high values causing primitives to shrink, therefore making raymarching more efficient, at the cost of quality.
 }
 \vspace{-2mm}
 \label{tab:overlap}
\end{table}
The influence of our primitive volume prior, $\mathcal{L}_{\text{vol}}$ in Section~\ref{sec:end_to_end_training}, is evaluated by training models with different weights, $\lambda_{\text{vol}} \in \{0.001, 0.01, 0.1\}$. We used models with 512 and 32,768 primitives in this evaluation (see Fig.~\ref{fig:results:boxprior} and Tab.~\ref{tab:overlap}). Larger weights lead to smaller scales and reduced overlap between adjacent primitives. This, in turn, leads to faster runtime performance since less overlap means that there are fewer primitives to evaluate at each marching step, as well as having less overall volume coverage. However, prior weights that are too large can lead to over shrinking, where holes start to appear in the reconstruction and image evidence is not sufficient to force them to expand.

\paragraph{Importance of the Opacity Fade Factor}
\begin{figure}[h]
    \centering
    \includegraphics[width=\linewidth]{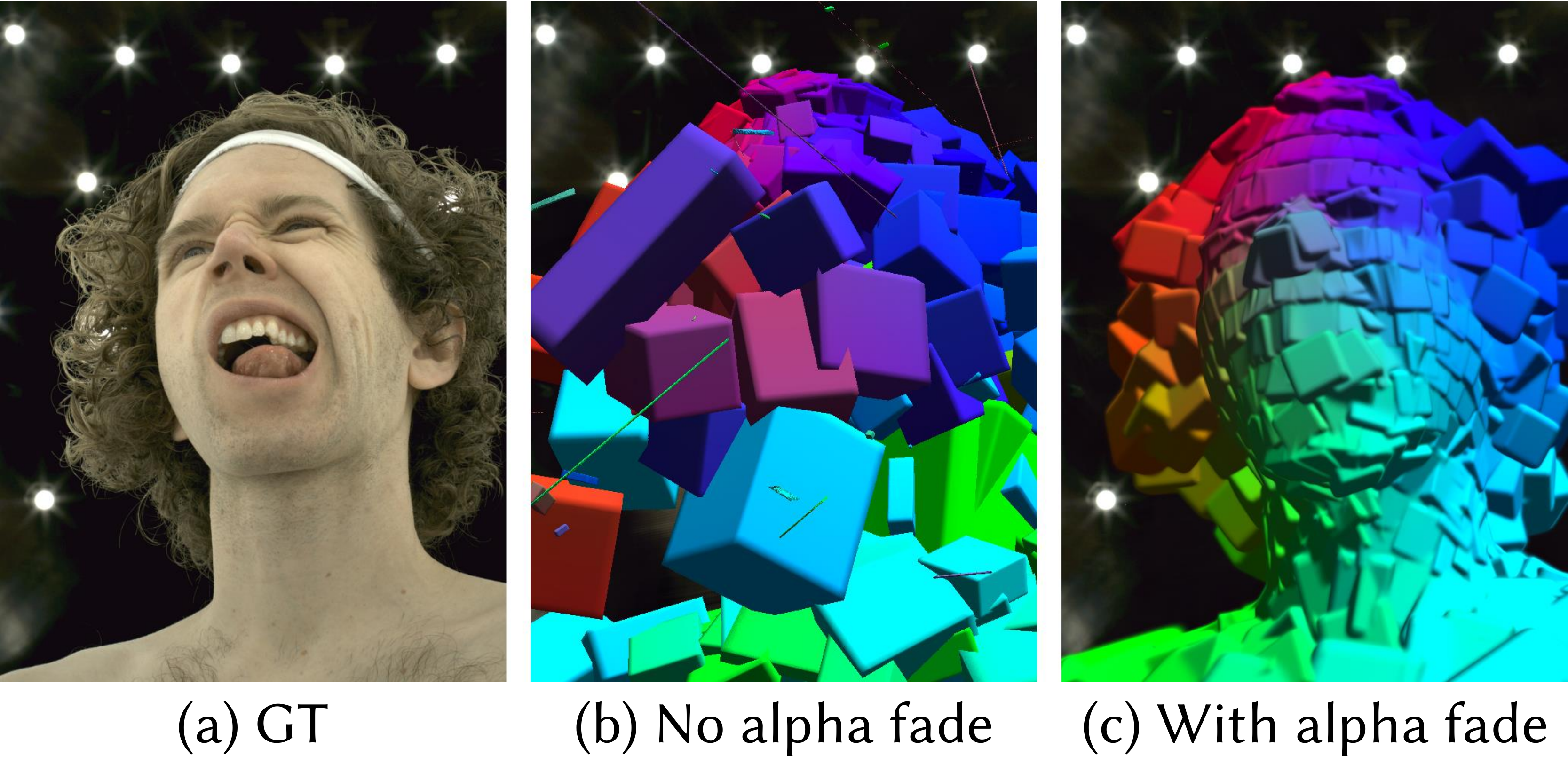}
    \vspace{-0.8cm}
    \caption{
    The opacity fade factor enables proper learning of primitive positions.
    We evaluate with and without opacity fade factor applied to the primitives.
    No opacity fade factor causes poor primitive positioning, drifting very far from the initialization, to a bad configuration with large primitives and high overlap because of poor gradients.
    }
    \label{fig:results:alphafade}
\end{figure}
We trained a model with 512 primitives with and without the opacity fade factor, $\mathbf{W}$, described in Section~\ref{sec:opacity_fade_factor}. As shown in Fig.~\ref{fig:results:alphafade}, opacity fade is critical for the primitives to converge to good configurations, as it allows gradients to properly flow from image to primitive position, rotation, and scaling. Without opacity fade, the gradients do not account for the impact of movement in the image plane at primitive silhouette edges. 
This results in suboptimal box configurations with large overlaps and coverage of empty space.
See the supplemental document for a quantitative comparison of the opacity fade factor.

\paragraph{Impact of Voxel Count and Raymarching Step Size}
\begin{figure}[t]
    \centering
    \includegraphics[width=\linewidth]{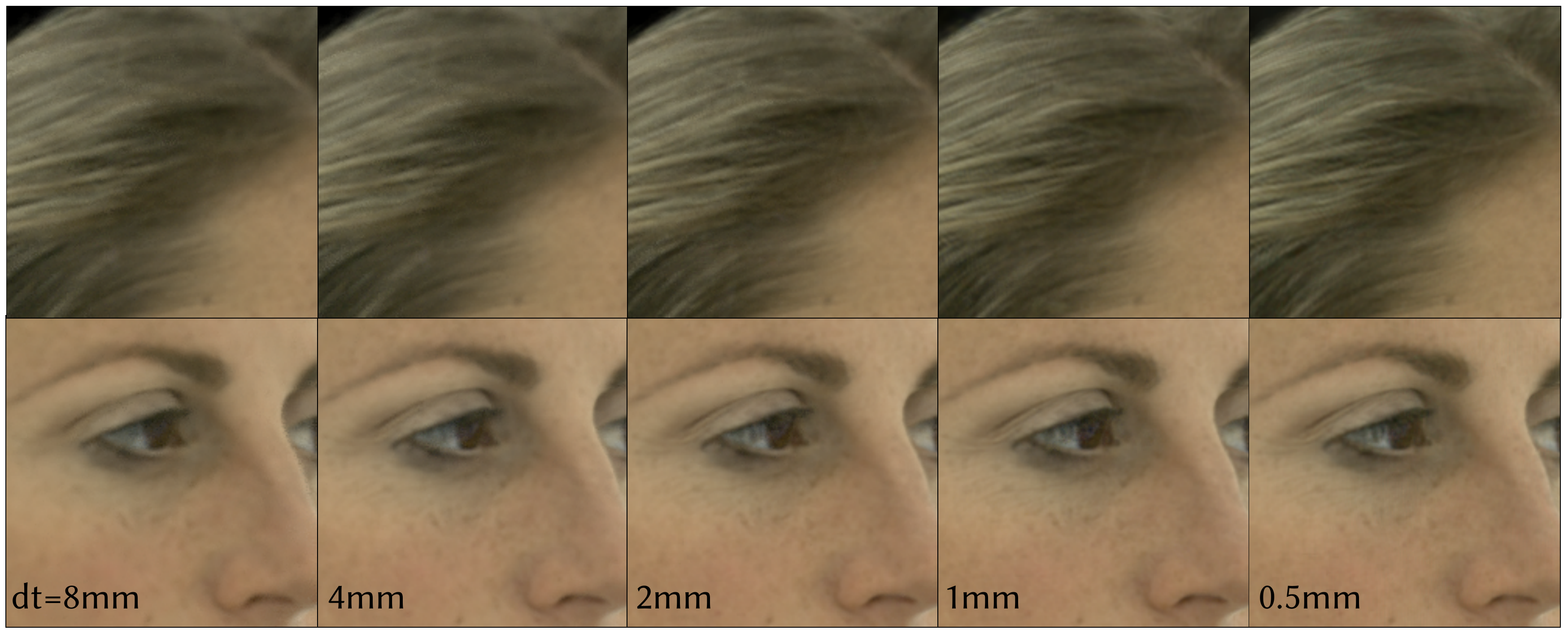}
    \vspace{-0.8cm}
    \caption{
    Here we show single-frame models of 16M voxels. Each model is trained and tested at the stepsize shown in the insets. Small step sizes are able to recover more detail, and reduce noise associated with raymarching at the cost of increased raymarching time.
    }
    \label{fig:results:stepsize}
\end{figure}

\begin{figure}[t]
    \centering
    \includegraphics[width=\linewidth]{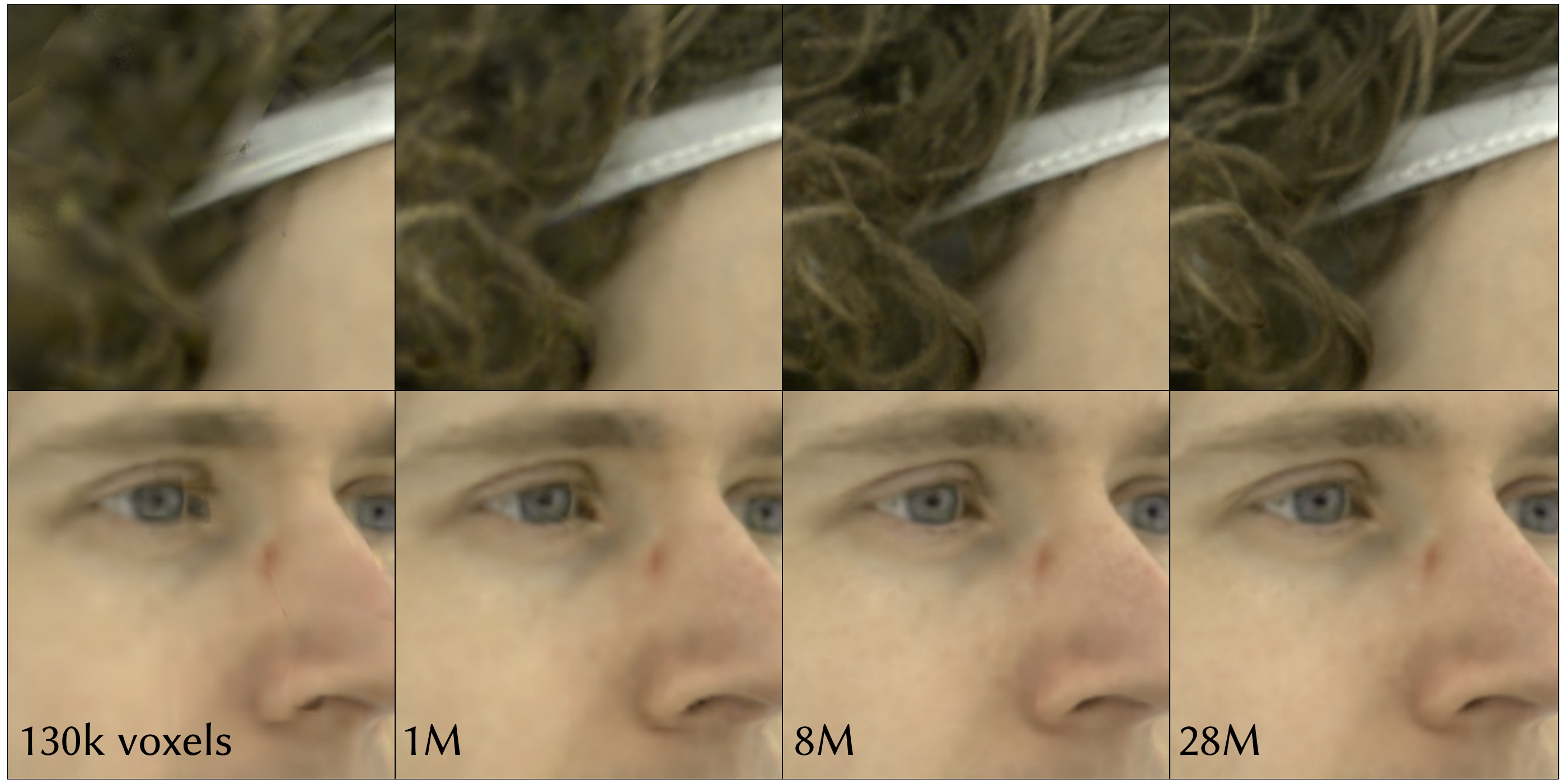}
    \vspace{-0.8cm}
    \caption{
    Effect of varying the number of voxels. Here we show a 256 primitive model at a stepsize of $1$mm, trained with $8^3$, $16^3$, $32^3$, and $48^3$ voxels per primitive, yielding the total number of voxels shown in the insets.
    Voxel models of more than $8$M voxels are generally too slow to run in realtime on current hardware.
    }
    \label{fig:qual:numvoxels}
\end{figure}

Fig.~\ref{fig:results:stepsize} and Fig.~\ref{fig:qual:numvoxels} illustrate the effects of different voxel counts and raymarching step sizes on perceived resolution (see the supplemental document for the quantitative impact of step size). Here, the same step size is used both during training and evaluation. Smaller marching step sizes recover more detail, such as hair and wrinkles, and result in lower reconstruction error on held out views. Likewise, more voxels provide sharper results. These gains in accuracy are, however, attained at the cost of performance. Decoding scales linearly with the number of voxels decoded, and raymarching scales linearly with step size.

\paragraph{Motion Model Architecture}
\begin{figure*}[h]
    \centering
    \includegraphics[width=\linewidth]{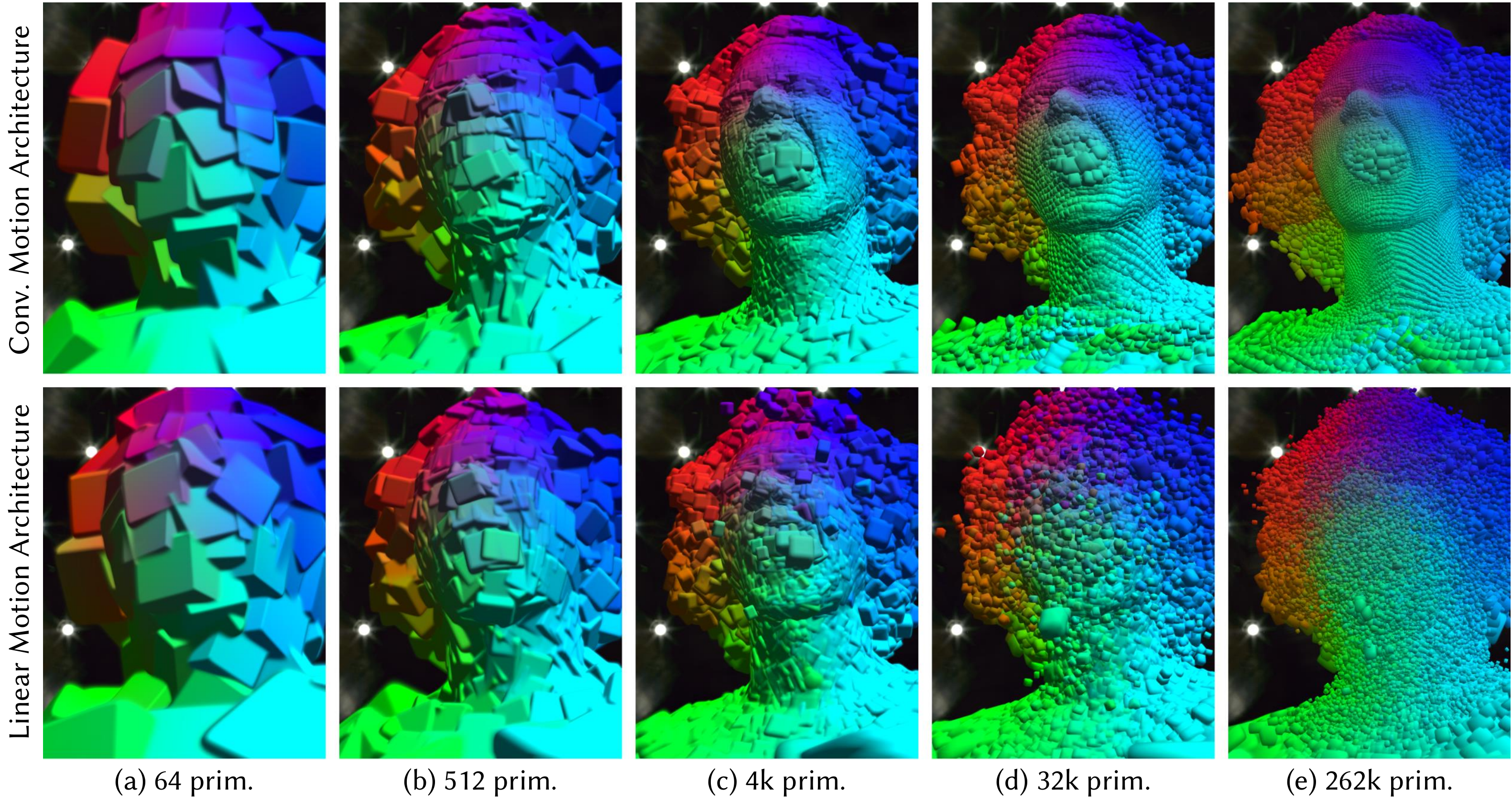}
    \vspace{-0.8cm}
    \caption{
    We evaluate two different motion models for the primitives.
    One uses a stack of convolutions, where each output pixel contains 9 channels representing the scaling, rotation, and translation of the corresponding primitive.
    The other uses a linear layer from the encoding to produce $9N_\text{boxes}$ channels that encode the scale, rotation, and translation of all primitives.
    The convolutional motion model produces boxes that closely follow the underlying surface and, as shown in Tab.~\ref{tab:motionmodel}, results in better reconstructions.
    }
    \label{fig:results:deconvmotion}
\end{figure*}
\begin{table*}[h]
\centering
 \begin{tabular}{r | c c | c c | c c | c c | c c}
 & \shortstack{64 prim. \\ conv} & \shortstack{64 prim \\ linear}
 & \shortstack{512 prim. \\ conv} & \shortstack{512 prim. \\ linear}
 & \shortstack{4k prim. \\ conv} & \shortstack{4k prim. \\ linear}
 & \shortstack{32k prim. \\ conv} & \shortstack{32k prim. \\ linear}
 & \shortstack{262k prim. \\ conv} & \shortstack{262k prim. \\ linear} \\
 \hline
 MSE ($\downarrow$)   & \cellcolor[rgb]{0.97,0.93,0.19}  42.2567
 & \cellcolor[rgb]{1.00,0.85,0.07}  43.7077
 & \cellcolor[rgb]{0.96,0.94,0.24}  41.9601
 & \cellcolor[rgb]{1.00,0.78,0.03}  44.7273
 & \cellcolor[rgb]{0.89,1.00,0.48}  \textbf{40.3211}
 & \cellcolor[rgb]{0.94,0.96,0.32}  41.3797
 & \cellcolor[rgb]{0.91,0.99,0.43}  40.6524
 & \cellcolor[rgb]{0.99,0.92,0.14}  42.5864
 & \cellcolor[rgb]{1.00,0.87,0.08}  43.4307
 & \cellcolor[rgb]{0.99,0.50,0.00}  50.4797
\\
 PSNR ($\uparrow$)   & \cellcolor[rgb]{0.98,0.93,0.17}  31.8718
 & \cellcolor[rgb]{1.00,0.84,0.06}  31.7252
 & \cellcolor[rgb]{0.97,0.94,0.21}  31.9024
 & \cellcolor[rgb]{1.00,0.77,0.02}  31.6251
 & \cellcolor[rgb]{0.89,1.00,0.48}  \textbf{32.0755}
 & \cellcolor[rgb]{0.94,0.96,0.31}  31.9629
 & \cellcolor[rgb]{0.91,0.99,0.43}  32.0399
 & \cellcolor[rgb]{1.00,0.91,0.11}  31.8381
 & \cellcolor[rgb]{1.00,0.86,0.07}  31.7528
 & \cellcolor[rgb]{0.99,0.50,0.00}  31.0996
\\
 SSIM ($\uparrow$)   & \cellcolor[rgb]{0.97,0.94,0.22}  0.9336
 & \cellcolor[rgb]{0.97,0.94,0.22}  0.9336
 & \cellcolor[rgb]{0.98,0.93,0.18}  0.9333
 & \cellcolor[rgb]{0.93,0.97,0.34}  0.9344
 & \cellcolor[rgb]{0.90,0.99,0.45}  0.9352
 & \cellcolor[rgb]{0.89,1.00,0.48}  \textbf{0.9354}
 & \cellcolor[rgb]{0.91,0.98,0.41}  0.9349
 & \cellcolor[rgb]{1.00,0.80,0.04}  0.9311
 & \cellcolor[rgb]{1.00,0.79,0.03}  0.9308
 & \cellcolor[rgb]{0.99,0.50,0.00}  0.9248
\\
 LPIPS ($\downarrow$)   & \cellcolor[rgb]{1.00,0.72,0.00}  0.2879
 & \cellcolor[rgb]{1.00,0.64,0.00}  0.2953
 & \cellcolor[rgb]{1.00,0.91,0.12}  0.2764
 & \cellcolor[rgb]{1.00,0.77,0.02}  0.2845
 & \cellcolor[rgb]{0.99,0.92,0.15}  0.2755
 & \cellcolor[rgb]{1.00,0.71,0.00}  0.2890
 & \cellcolor[rgb]{0.89,1.00,0.48}  \textbf{0.2670}
 & \cellcolor[rgb]{1.00,0.73,0.00}  0.2877
 & \cellcolor[rgb]{1.00,0.91,0.10}  0.2767
 & \cellcolor[rgb]{0.99,0.50,0.00}  0.3059
\\
\hline
 decode ($\downarrow$)   & \cellcolor[rgb]{0.99,0.50,0.00}  57.5364
 & \cellcolor[rgb]{0.99,0.51,0.00}  56.9789
 & \cellcolor[rgb]{0.99,0.55,0.00}  54.0634
 & \cellcolor[rgb]{0.99,0.56,0.00}  53.6730
 & \cellcolor[rgb]{1.00,0.78,0.02}  39.3384
 & \cellcolor[rgb]{1.00,0.81,0.04}  37.9490
 & \cellcolor[rgb]{1.00,0.77,0.02}  39.6899
 & \cellcolor[rgb]{1.00,0.84,0.06}  36.5781
 & \cellcolor[rgb]{0.91,0.99,0.42}  26.4612
 & \cellcolor[rgb]{0.89,1.00,0.48}  \textbf{25.0964}
\\
 raymarch ($\downarrow$)   & \cellcolor[rgb]{0.89,1.00,0.48}  \textbf{8.7660}
 & \cellcolor[rgb]{0.89,1.00,0.48}  9.1570
 & \cellcolor[rgb]{0.90,1.00,0.47}  10.6198
 & \cellcolor[rgb]{0.91,0.99,0.43}  14.7531
 & \cellcolor[rgb]{0.90,0.99,0.44}  13.3397
 & \cellcolor[rgb]{0.95,0.95,0.28}  32.1545
 & \cellcolor[rgb]{0.92,0.98,0.38}  19.8951
 & \cellcolor[rgb]{1.00,0.83,0.06}  72.9552
 & \cellcolor[rgb]{0.96,0.94,0.25}  35.8098
 & \cellcolor[rgb]{0.99,0.50,0.00}  184.4986
\\
 total ($\downarrow$)   & \cellcolor[rgb]{0.93,0.97,0.35}  66.3024
 & \cellcolor[rgb]{0.93,0.97,0.35}  66.1359
 & \cellcolor[rgb]{0.93,0.97,0.37}  64.6831
 & \cellcolor[rgb]{0.94,0.96,0.33}  68.4260
 & \cellcolor[rgb]{0.89,1.00,0.48}  \textbf{52.6781}
 & \cellcolor[rgb]{0.94,0.96,0.31}  70.1036
 & \cellcolor[rgb]{0.91,0.98,0.41}  59.5850
 & \cellcolor[rgb]{1.00,0.84,0.06}  109.5333
 & \cellcolor[rgb]{0.92,0.98,0.39}  62.2710
 & \cellcolor[rgb]{0.99,0.50,0.00}  209.5950
\\
 \end{tabular}
 \caption{
 Quantitative evaluation of convolutional motion model vs. linear motion model. The convolutional architecture for producing primitive location, orientation, and size results in primitives that are more coherent with the mesh surface, which allows it to achieve higher quality and better performance than a linear model.
 }
 \vspace{-2mm}
 \label{tab:motionmodel}
\end{table*}
The motion model, $\mathcal{D_{\text{pRs}}}$, regresses position, rotation, and scale deviations of the volumetric primitive from the underlying guide mesh; $\{\boldsymbol{\delta}_{{t}_k}, \boldsymbol{\delta}_{{R}_k}, \boldsymbol{\delta_{\mathbf{s}_k}}\}_{k=1}^{N_{\text{prim}}}$. 
We compared the convolutional architecture we proposed in Section \ref{sec:method} against a simpler linear layer that transforms the latent code $\mathbf{z}$ to a $9N_{\text{prim}}$-dimensional vector, comprising the stacked motion vectors, i.e., three dimensions each for translation, rotation and scale.
Tab.~\ref{tab:motionmodel} shows that our convolutional architecture outperforms the simple linear model for almost all primitive counts, $N_{\text{prim}}$. A visualization of their differences can be seen in Fig.~\ref{fig:results:deconvmotion}, where our convolutional model produces primitive configurations that follow surfaces in the scene more closely than the linear model, which produces many more "fly away" zero-opacity primitives, wasting resolution.

\subsection{Comparisons}
We compared MVP against the current state of the art in neural rendering for both static and dynamic scenes. As our qualitative and quantitative results show, we outperform the current state of the art in reconstruction quality as well as runtime performance.

\paragraph{Neural Volumes}
\begin{figure*}[h]
    \centering
    \includegraphics[width=\linewidth]{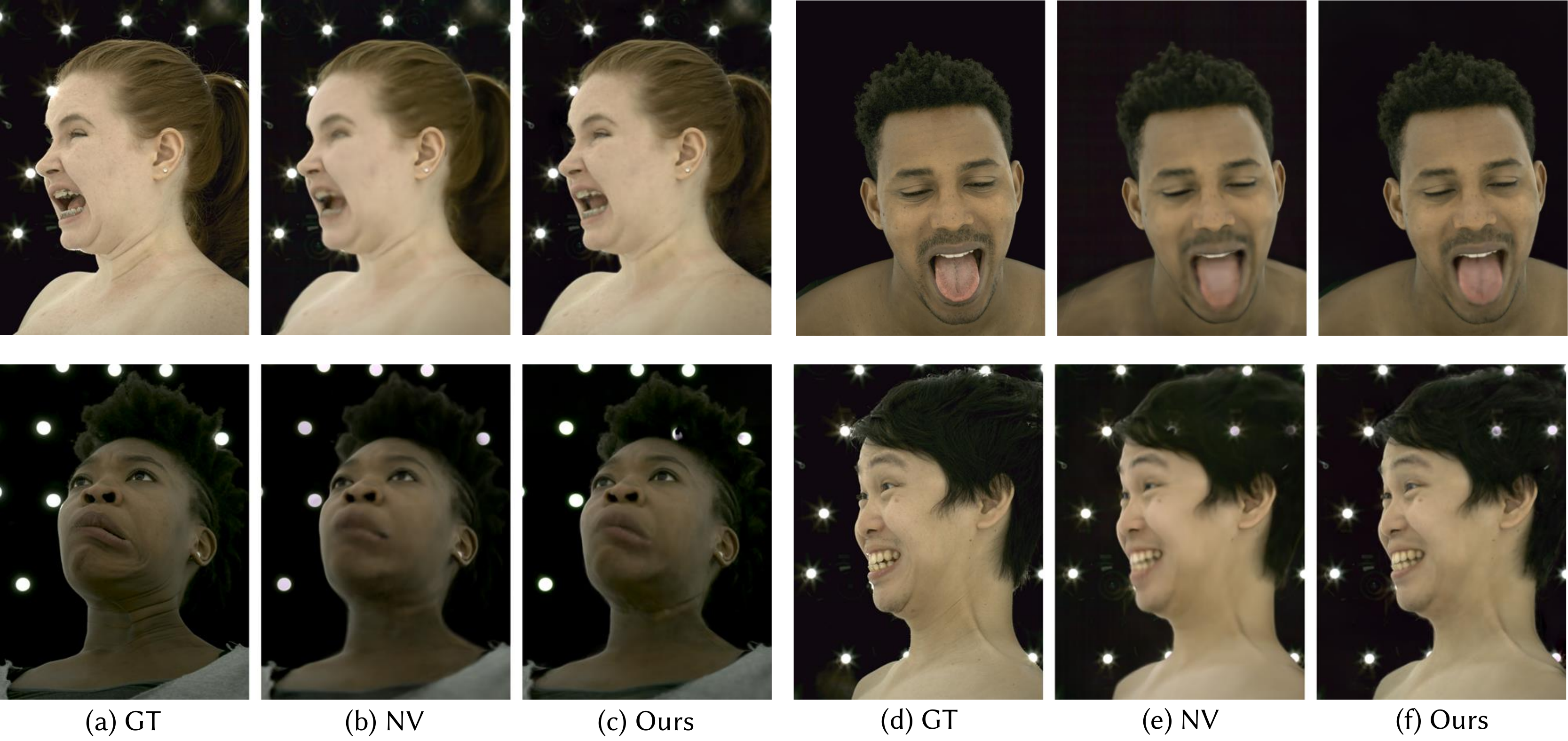}
    \vspace{-0.8cm}
    \caption{
    We compare with Neural Volumes for several additional subjects. In all cases, our method can better reproduce facial expressions with more accuracy and finer detail.
    }
    \label{fig:results:subjects}
\end{figure*}
We compare to Neural Volumes (NV) \cite{lombardi2019nv} on several challenging dynamic sequences, see Fig.~\ref{fig:results:subjects}.
Our approach obtains sharper and more detailed reconstructions, while being much faster to render.
We attribute this to our novel mixture of volumetric primitives that can concentrate representation resolution and compute in occupied regions in the scene and skip empty space during raymarching. 
Quantitative comparisons are presented in Tab.~\ref{tab:boxes}.
As can be seen, we outperform Neural Volumes (NV) \cite{lombardi2019nv} in terms of SSIM and LPIPS.

\paragraph{Neural Radiance Fields (NeRF)}
\begin{figure}[h]
    \centering
    \includegraphics[width=\linewidth]{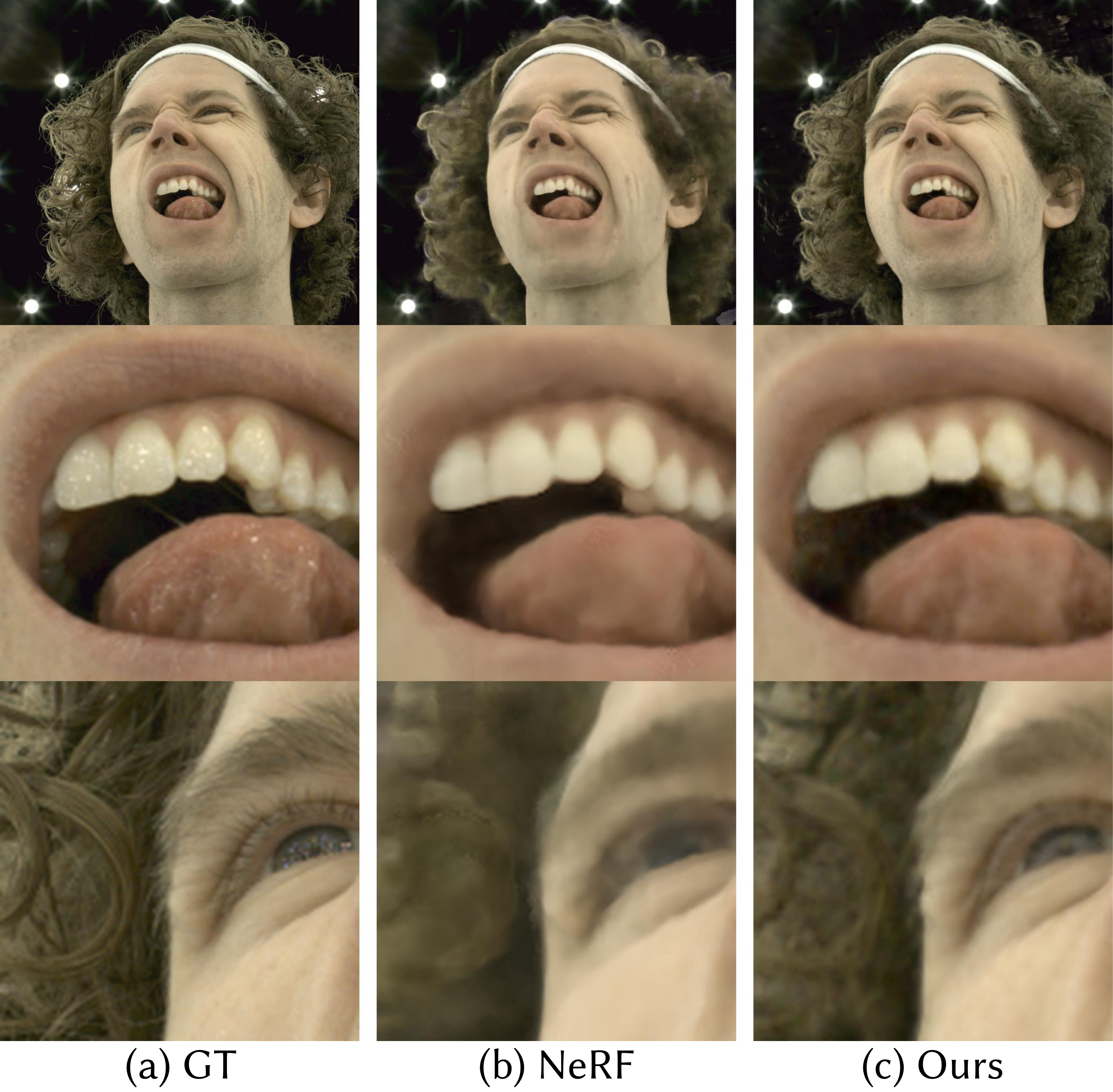}
    \vspace{-0.8cm}
    \caption{
    Comparison to Neural Radiance Fields (NeRF) \cite{mildenhall2019llff}.
    NeRF is a novel view synthesis approach for static scenes.
    We compare by training both models for a single time step.
    NeRF excels at representing geometric detail, as can be seen in the teeth, but struggles with planar texture detail.
    Our approach obtains more detailed results, while being orders of magnitude faster to render.
    }
    \label{fig:results:nerf}
    \vspace{-12pt}
\end{figure}
\begin{table}[t]
\centering
 \begin{tabular}{r | c c c }
 & \shortstack{NeRF} & \shortstack{Ours \\(single frame)} & \shortstack{Ours \\(multi-frame)} \\
 \hline
 MSE ($\downarrow$)   & \cellcolor[rgb]{0.99,0.50,0.00}  47.6849 & \cellcolor[rgb]{0.89,1.00,0.48}  \textbf{37.9261} & \cellcolor[rgb]{0.93,0.97,0.34}  38.8223\\
 PSNR ($\uparrow$)   & \cellcolor[rgb]{0.99,0.50,0.00}  31.3470 & \cellcolor[rgb]{0.89,1.00,0.48}  \textbf{32.3414} & \cellcolor[rgb]{0.94,0.96,0.32}  32.2400\\
 SSIM ($\uparrow$)   & \cellcolor[rgb]{0.99,0.50,0.00}  0.9279 & \cellcolor[rgb]{0.89,1.00,0.48}  \textbf{0.9381} & \cellcolor[rgb]{0.99,0.92,0.15}  0.9359\\
 LPIPS ($\downarrow$)   & \cellcolor[rgb]{0.99,0.50,0.00}  0.3018 & \cellcolor[rgb]{0.89,1.00,0.48}  \textbf{0.2080} & \cellcolor[rgb]{1.00,0.76,0.01}  0.2529\\
\hline
 decode ($\downarrow$)   & -  & \cellcolor[rgb]{0.89,1.00,0.48}  \textbf{18.8364} & \cellcolor[rgb]{0.99,0.50,0.00}  19.4213\\
 raymarch ($\downarrow$)   & -  & \cellcolor[rgb]{0.89,1.00,0.48}  \textbf{31.5097} & \cellcolor[rgb]{0.99,0.50,0.00}  37.1573\\
 total ($\downarrow$)   & \cellcolor[rgb]{0.99,0.50,0.00}  27910.0000 & \cellcolor[rgb]{0.89,1.00,0.48}  \textbf{50.3462} & \cellcolor[rgb]{0.89,1.00,0.48}  56.5786\\
 \end{tabular}
 \caption{
 Comparison of our approach to NeRF trained on both a single frame and a sequence (here evaluated only on a static frame).
 Our approach improves over NeRF in performance by 1000x, and improves quality even when trained on a long sequence.
 }
 \vspace{-2mm}
 \label{tab:nerfcomp}
\end{table}
We also compare our approach to a Neural Radiance Field (NeRF) \cite{mildenhall2020nerf}, see Fig.~\ref{fig:results:nerf}.
Since NeRF is an approach for novel view synthesis of static scenes, we trained it using only a single frame of our capture sequence.
We compared it against MVP trained on both the same frame and the entire sequence of approximately 20,000-frame.
A visualization of the differences between NeRF and MVP on the static frame is shown in Fig.~\ref{fig:results:nerf}. NeRF excels at representing geometric detail, as can be seen in the teeth, but struggles with planar texture detail, like the texture on the lips or eyebrows. MVP captures both geometric and texture details well. 
Quantitative results comparing the methods is presented in Tab.~\ref{tab:nerfcomp}, where our method outperforms NeRF on all metrics, even when trained using multiple frames. 
Finally, our approach improves over NeRF in runtime performance by three orders of magnitude. 

\section{Limitations}
We have demonstrated high quality neural rendering results for dynamic scenes.
Nevertheless, our approach is subject to a few limitations that can be addressed in follow-up work.
First, we require a coarse tracked mesh to initialize the positions, rotations, and scale of the volumetric primitives.
In the future, we'd like to obviate this requirement and allow the primitives to self-organize based only on the camera images.
Second, the approach requires a high-end computer and graphics card to achieve real-time performance.
One reason for this is the often high overlap between adjacent volumetric primitives.
Thus, we have to perform multiple trilinear interpolations per sample point, which negatively impacts runtime.
It would be interesting to incorporate regularization strategies to minimize overlap, which could lead to a significant performance boost.
Finally, the number of employed primitives is predefined and has to be empirically determined for each scene type.
It is an interesting direction for future work to incorporate this selection process into the optimization, such that the best setting can be automatically determined.
Despite these limitations, we believe that our approach is already a significant step forward for real-time neural rendering of dynamic scenes at high resolutions.

\section{Conclusion}
We have presented a novel 3D neural scene representation that handles dynamic scenes, is fast to render, drivable, and can represent 3D space at high resolution.
At the core of our scene representation is a novel mixture of volumetric primitives that is regressed by an encoder-decoder network.
We train our representation based on a combination of 2D and 3D supervision.
Our approach generalizes volumetric and primitive-based paradigms under a unified representation and combines their advantages, thus leading to high performance decoding and efficient rendering of dynamic scenes.
As our comparisons demonstrate, we obtain higher quality results than the current state of the art.
We hope that our approach will be a stepping stone towards highly efficient neural rendering approaches for dynamic scenes and that it will inspire follow-up work.

\bibliographystyle{ACM-Reference-Format}
\typeout{}
\bibliography{egbib}

\end{document}